**Validation of quasi-two-dimensional model of convection in a transverse magnetic field**

Alexander Gelfgat
*School of Mechanical Engineering, Faculty of Engineering, Tel-Aviv University, Ramat Aviv, Tel-Aviv 6997801, Israel*

## Abstract

In this study we examine applicability of the quasi-two-dimensional (Q2D) model of natural convection in a laterally heated rectangular box subject to a transverse magnetic field. For this purpose, results obtained using the Q2D model are compared with fully three-dimensional calculations for rectangular boxes with a square cross-section and horizontally elongated ones with the length to height ratio 8. The width to height ratio is varied from 1 to 10. It is argued that the Q2D model is valid under certain conditions, which include perfect thermal and electrical insulation of the spanwise boundaries and reflection symmetry with respect to the spanwise midplane. It is shown that under these conditions, the Q2D model yields quantitatively correct results for boxes with square cross-section, as well as for horizontally elongated cavities, independently on the width to height ratio. Three-dimensional results obtained for boxes with perfectly conducting walls exhibit asymptotics at gradually increasing width ratio, which, however, does not agree with the predictions of an amended Q2D model. It is shown also that when is applicable, the Q2D model can correctly predict steady-oscillatory transition of three-dimensional flows. It is proposed to seek for extensions of the Q2D model for cavities with perfectly conducting spanwise boundaries.

## 1. Introduction

The quasi-two-dimensional (Q2D) model was developed at first for MHD flows in rectangular ducts under effect of transverse magnetic field [1-4]. Later, in [5-7] it was extended to natural convection flows, when the magnetic field is orthogonal to the main circulation, i.e., is transverse, so that it does not interact with the flow in a purely 2D configuration, however, its damping effect on the 3D convection is well documented in several experimental and numerical studies of three-dimensional models (see [87-1420] and references therein). Convective flows in the transverse magnetic field were studied experimentally [7-13], as well as numerically, using either fully 3D [7,14-20] or Q2D [5,2422-24] formulation. Studies [5,1210-141316] addressed onset of convection and supercritical flows in the Rayleigh-Bénard configuration with heating form below. Other case of interest is convection in cavities with heating from the side [6,7,17-24]. In most of these studies [109-1314,17-20] the quasi-two-dimensional flow patterns were observed, thus showing relevance of the Q2D model.

As explained in [7], the experimental validation of the Q2D model is quite difficult, because large thermal conductivity of liquid metals alters uniform temperatures of the boundaries, which, in the idealistic case, are assumed to be isothermal. This can be a reason why quasi-two-dimensional structure observed in the experiments were not directly compared with predictions of the Q2D model. The agreement between fully 3D numerical results and those yielded by the Q2D model was examined quite rarely [6,24], and, in the opinion of the author, deserves a more detail study. In the study [76] the authors found that Q2D model reproduces the number of convective rolls, but does not reproduce the Nusselt number. In the study [1924] the quantitative comparison was done only for the Nusselt number, while other results exhibited only qualitative agreements. A good agreement between the Q2D flow patterns and those of fully 3D calculations was reported in [19].

The purpose of this study is a more rigorous comparison of numerical solutions obtained for the Q2D model and by the fully three-dimensional computations. We recall the evaluation of the Q2D model for convective flow in a box, and discuss all assumptions needed to make the model valid. We show that besides the requirement of sufficiently strong magnetic field, the spanwise boundaries must be electrically insulated, and the problem must be symmetric with respect to the spanwise midplane. For the 3D boxes with the unity aspect ratio (=length/height), we show that when these conditions are satisfied, there is a very good agreement between the integral flow characteristics, i.e., the Nusselt number and the total kinetic energy, as well as between profiles of the Q2D and spanwise averaged 3D flows. This good agreement holds for the gradually increased width ratio (width/height), which was varied between 1 and 10. This agreement is observed even when the three-dimensional flow bifurcates to periodic or chaotic oscillatory regime, so that the comparison can be done only for the time-averaged flow characteristics.

We discuss also how the Q2D model can be extended for perfectly electrically conducting spanwise boundaries. However, similarly to the results of [22], the model we obtain does not yield correct quantitative results. At the same time, it does exhibit asymptotics at large width ratios, which agree with the Q2D predictions only when the flows at large Hartmann numbers are subject to a very strong electromagnetic damping. At the same time, existing of asymptotics at moderate Hartmann numbers shows that a simplified two-dimensional model can be possibly developed also for this case.

Finally, we consider two examples of steady – oscillatory transition in fully 3D and Q2D models, and show that the Q2D model can quite accurately predict critical Grashof number, frequency of oscillations, and spanwise-averaged distribution of oscillations amplitude.

## 2. Formulation of the problem

### *2.1. Full three-dimensional problem*

We consider convective flow in a three-dimensional rectangular box in the presence of a horizontal magnetic field $\mathbf{B} = B_0\mathbf{e}_y$ as shown in Fig. 1a. The height, length, and width of the box are $H$, $L$, and $D$, respectively. The walls of the box at $x = 0, L$ are kept at constant values of temperature, $T_{hot}$ and $T_{cold}$, respectively, while those at $z = 0, H$ and $y = -\frac{D}{2}, \frac{D}{2}$ are thermally insulating. The fluid has electrical conductivity $\sigma$, kinematic viscosity $\nu$, thermal diffusivity $\kappa$, and density, $\rho = \rho_0[1 - \beta(T - T_{cold})]$, where $\beta$ is the thermal expansion coefficient, and $\rho_0$ is the fluid density at temperature $T_{cold}$.

To render equations dimensionless we choose $D^2/\nu, \nu/D, \rho\,\nu^2/D^2$ as scales of the length, time $t$, the velocity $\boldsymbol{v} = (u, v, w)$ and the pressure $p$, respectively, where $\nu$ is the fluid kinematic viscosity and $\rho$ is the density. The temperature is rescaled to a dimensionless function by $T \rightarrow (T - T_{cold})/(T_{hot} - T_{cold})$. Additionally, the dimensionless time, velocity and pressure are scaled, respectively, by $Gr^{-1/2}$, $Gr^{1/2}$, and $Gr$, where $Gr = g\beta(T_{hot} - T_{cold})H^3/\nu^2$ is the Grashof number, $g$ is the acceleration due to gravity, and $\beta$ is the thermal expansion coefficient. The electric current and electric potential are scaled by $Gr^{-1/2}\sigma\nu B_0/D$ and $Gr^{-1/2}\nu B_0$, respectively. Assuming that the magnetic Prandtl number is very small, the dimensionless governing equations in the Boussinesq approximation read

$$\frac{\partial \boldsymbol{v}}{\partial t} + (\boldsymbol{v} \cdot \nabla)\boldsymbol{v} = -\nabla p + \frac{1}{Gr^{1/2}} \Delta \boldsymbol{v} + T\boldsymbol{e}_z + \frac{Ha^2}{Gr} \boldsymbol{j} \times \boldsymbol{e}_y, \quad (1)$$

$$\boldsymbol{j} = -\nabla \phi + \boldsymbol{v} \times \boldsymbol{e}_y, \quad (2)$$

$$\nabla \cdot \boldsymbol{v} = 0, \quad \nabla \cdot \boldsymbol{j} = 0 \quad (3,4)$$

$$\frac{\partial T}{\partial t} + (v \cdot \nabla)T = \frac{1}{PrGr^{1/2}} \Delta T \quad (5)$$

where $Ha = B_0 H(\sigma/\rho_0 \nu)^{1/2}$ , and $Pr = \nu/\kappa$, are the Hartmann, and Prandtl numbers, respectively.

The boundaries are no-slip

$$\boldsymbol{v} = 0 \quad \text{at all the walls.} \quad (6)$$

For the following, we restrict ourselves to the thermally insulated horizontal and spanwise walls, so that the thermal boundary conditions are:

$$T = 1 \quad \text{at} \quad x = 0, \quad (7)$$

$$T = 0 \quad \text{at} \quad x = A, \quad (8)$$

$$\partial T/\partial z = 0 \quad \text{at} \quad z = 0,1, \quad (9)$$

$$\partial T/\partial y = 0 \quad \text{at} \; y = \pm \frac{W}{2} \quad . \quad (10)$$

In the above, $A = L/H$ and $W = D/H$ are aspect and width ratios of the box, respectively.

Further, we consider the vertical and horizontal walls either perfectly electrically insulating or perfectly electrically conducting. In the first case the normal component of electric current vanishes at the boundaries. Taking into account the no-slip conditions for the velocity and eq. (2), it reads

$$j_x = \frac{\partial \phi}{\partial x} = 0 \quad \text{at} \quad x = 0, A, \quad (11a)$$

$$j_y = \frac{\partial \phi}{\partial y} = 0 \quad \text{at} \quad y = \pm \frac{W}{2}, \quad (11b)$$

$$j_z = \frac{\partial \phi}{\partial z} = 0 \quad \text{at} \quad z = 0,1 \quad \text{for the perfectly electrically insulating walls.} \quad (11c)$$

In the case of electrically conducting boundaries we consider a model of thin conducting wall, which leads to the boundary condition [25]:

$$j_x = -c\left(\frac{\partial^2 \phi}{\partial y^2} + \frac{\partial^2 \phi}{\partial z^2}\right) \quad \text{at} \quad x = 0, A, \quad (12a)$$

$$j_y = -c\left(\frac{\partial^2 \phi}{\partial x^2} + \frac{\partial^2 \phi}{\partial z^2}\right) \quad \text{at} \quad y = \pm \frac{W}{2}, \quad (12b)$$

$$j_z = -c\left(\frac{\partial^2\phi}{\partial x^2}+\frac{\partial^2\phi}{\partial y^2}\right) \quad \text{at} \quad z = 0,1 \quad . \tag{12c}$$

Here $c = d\sigma_w/\sigma$, $\sigma_w$is the wall conductivity, and $d$ is the wall dimensionless thickness. We are interested in two limiting cases, when $c = 0$ for the perfectly insulating wall, and $c = \infty$ for the perfectly conducting one. Note, that for the last case the current density at the boundaries will be finite only if two-dimensional Laplacians in the boundary conditions (12a-c) vanish at the boundaries. Assuming all the boundaries to be equipotential in this case, we prescribe the zero value of the electric potential. If, on the other hand, $c = 0$, the equation (2) will result in the Neumann boundary conditions (11) for the electric potential. For the folowing, we assume that the electro-physical properties of the boundaries allow for the symmetry with respect to the midplane $y = 0$.

*2.2 Quasi-two-dimensional (Q2D) model*

The quasi two-dimensional model is based on the averaging in *y*-direction. For an arbitrary function $F(x, y, z, t)$ the averaging is defined by

$$\bar{F}(x,z,t) = \frac{1}{W}\int_0^W F(x,y,z,t)dy \; . \tag{13}$$

Before proceeding to averaging of the governing equations, some comments with connection to the considered flow model are needed. The reflection symmetry of the problem with respect to the plane $y = 0$ is $\{v_x, v_y, v_z, \theta\}(x, y, z) = \{v_x, -v_y, v_z, \theta\}(x, -y, z)$ [28]. Thus, the component $v_y$, is an odd function of $y$, so that $\bar{v}_y = 0$. Therefore, the *y*-averaged velocity is a two-dimensional vector $\bar{\boldsymbol{v}} = (\bar{v}_x, 0, \bar{v}_z)$. Due to the no-slip conditions at the spanwise boundaries, also $\overline{\partial v_y/\partial y} = v_y(W/2) - v_y(-W/2) = 0$. Therefore, the averaged continuity equation (3) results analytically into

$$\overline{\nabla\cdot\mathbf{v}} = \nabla_\perp\cdot\bar{\mathbf{v}} = \frac{\partial\bar{v}_x}{\partial x}+\frac{\partial\bar{v}_z}{\partial z} = 0\,, \quad \nabla_\perp = \boldsymbol{e}_x\frac{\partial}{\partial x}+\boldsymbol{e}_y\frac{\partial}{\partial z}\;, \tag{14}$$

and the *y*-averaged velocity is divergence-free. Now, averaging equation (4), and using boundary conditions (12), we obtain

$$\overline{\nabla\cdot\boldsymbol{J}} = \frac{\partial\bar{J}_x}{\partial x}+\frac{\partial\bar{J}_z}{\partial x}+j_y\left(y=\frac{W}{2}\right)-j_y(y=-\frac{W}{2}) \tag{15}$$

In the general case the last two terms of Eq. (15) describe the effect of the non-zero spanwise wall conductivity, while at the electrically insulating spanwise boundaries these terms vanish. Taking into account the reflection symmetry with respect to the midplane $y = 0$ we observe that the electric potential, as well as *x*- and *z*-components of the electric current must be even functions of *y*, while the *y*-component of the current must be an odd function. The latter means that the difference of last two terms in Eq. (15) will not be zero if the spanwise borders are not perfectly insulating, and this non-zero value should be taken into account in the forthcoming derivations. Averaging Eq. (2), and assuming that all boundaries are equipotential, we obtain

$$\bar{\boldsymbol{j}} = -\nabla_\perp \bar{\phi} + \bar{\boldsymbol{v}} \times \boldsymbol{e}_y + \phi\left(y = -\tfrac{W}{2}\right) - \phi\left(y = \tfrac{W}{2}\right) = -\nabla_\perp \bar{\phi} + \bar{\boldsymbol{v}} \times \boldsymbol{e}_y, \tag{16}$$

and

$$\nabla_\perp \times \bar{\boldsymbol{j}} = -\nabla_\perp \times (\nabla_\perp \bar{\phi}) + \nabla_\perp \times (\bar{\boldsymbol{v}} \times \boldsymbol{e}_y) = 0 \ . \tag{17}$$

In the case of electrically insulating boundaries, Eqs. (15) and (17) imply that $\bar{\mathbf{j}}$ is a harmonic vector function in every (*x,z*) plane. Therefore, $\nabla_\perp^2 \bar{\boldsymbol{j}} = 0$, and owing to the boundary conditions (11), the *y*-averaged electric current vanishes everywhere, i.e., $\bar{\mathbf{j}} = \mathbf{0}$, as it was considered in [22-24]. In the case of electrically conducting boundaries, the averaged electric current is not zero, and more evaluations are needed (see below).

To average the momentum equation, we recall that due to the strong electromagnetic dumping, the velocity component $v_y$ is assumed to be much smaller than either $v_x$ or $v_z$, so that all the terms involving $v_y$ are neglected. Additionally we assume that variations of the components $v_x$ and $v_z$ along the coordinate *y* are similar and are defined by the same profile $h(y)$

$$v_x(x,y,z,t) = \bar{v}_x(x,z,t)h(y), \ \ v_z(x,y,z,t) = \bar{v}_z(x,z,t)h(y)\,, \tag{18}$$

$$h(y) = \frac{Ha}{Ha-2}\left\{1 - exp\left[Ha\left(y - \frac{W}{2}\right)\right] - exp\left[-Ha\left(y + \frac{W}{2}\right)\right]\right\}. \tag{19}$$

The profile $h(y)$ describes an exponential decay of velocity from the midplane $y = 0$ towards the boundaries $y = \pm W/2$. It is an even function of $y$, so that $\overline{h'} = 0$, and

$$\bar{h} = \frac{1}{W}\int_{-W/2}^{W/2} h(y)dy = \frac{HaW - 2\left(1 - e^{-HaW}\right)}{(Ha-2)W} \xrightarrow[Ha\to\infty]{} 1 \ . \tag{20}$$

Examples of this profile at different Hartmann numbers are shown in Fig. 2. Using the above assumption, we derive the *y*-averaged non-linear terms of the momentum equation, which reduces to evaluation of

$$\overline{h^2} = \frac{1}{W}\int_{-W/2}^{W/2} h^2(y)dy = \frac{Ha\left(WHa + 2WHae^{-WHa} - 3 + 4e^{-WHa} - e^{-2WHa}\right)}{W(Ha-2)} \xrightarrow[Ha\to\infty]{} 1\,, \tag{21}$$

and at large *Ha*, assuming a negligibly small $v_y$,

$$\overline{(\boldsymbol{v}\cdot\nabla)\boldsymbol{v}} = (\bar{\boldsymbol{v}}\cdot\nabla_\perp)\bar{\boldsymbol{v}}\,. \tag{24}$$

For *y*-averaging of the dissipative term we evaluate

$$\overline{h''} = \frac{1}{W}\int_{-W/2}^{W/2} h''(y)dy = \frac{2Ha^2(e^{-HaW}-1)}{(Ha-2)W} \xrightarrow[Ha\to\infty]{} -\frac{2Ha}{W}\,, \tag{25}$$

and at large $Ha$

$$\overline{\Delta\boldsymbol{v}} = \Delta_\perp \boldsymbol{v} - \frac{2Ha}{W}\boldsymbol{v}, \qquad \Delta_\perp = \nabla_\perp^2\ , \tag{26}$$

Since $v_y$ is an odd function of $y$, all other terms of the $y$-component of the momentum equation, e.g., $\partial p/\partial y$, also must be odd functions, so that the $y$-average of this component is zero. Since $\partial p/\partial y$ is an odd function, $p$ can be considered as an even one, so that the derivatives $\partial p/\partial x$ and $\partial p/\partial z$ also are even functions, and their $y$-average values are $\partial \bar{p}/\partial x$ and $\partial \bar{p}/\partial z$. Finally, the $y$-averaged momentum equation (1) reads

$$\frac{\partial \bar{\boldsymbol{v}}}{\partial t} + (\bar{\boldsymbol{v}}\cdot\nabla_\perp)\bar{\boldsymbol{v}} = -\nabla_\perp \bar{p} + \frac{1}{Gr^{1/2}}\Delta_\perp \bar{\boldsymbol{v}} - \frac{Hd}{Gr^{1/2}}\bar{\boldsymbol{v}} + \bar{T}\boldsymbol{e}_z + \frac{Ha^2}{Gr}\bar{\boldsymbol{J}}\times\boldsymbol{e}_y, \tag{27}$$

where $Hd = 2Ha/W$ is the governing parameter of the Q2D model, and is different from the Hartmann number. It is stressed, however, that only two out of three parameters $Ha$, $Hd$, and $W$ are independent. This parameter was used in [22-23]. In studies [7,23], the width ratio was $W = 1$, and the governing parameter was $2Ha$, which agrees to the above derivations. To evaluate the last term of the above equations, we substitute Eq. (16) into Eq. (27) and arrive to

$$\frac{\partial \bar{\boldsymbol{v}}}{\partial t} + (\bar{\boldsymbol{v}}\cdot\nabla_\perp)\bar{\boldsymbol{v}} = -\nabla_\perp \bar{p} + \frac{1}{Gr^{1/2}}\Delta_\perp \bar{\boldsymbol{v}} - \frac{Hd}{Gr^{1/2}}d\bar{\boldsymbol{v}} + \bar{T}\boldsymbol{e}_z + \frac{Ha^2}{Gr}\left(-\nabla\bar{\phi} + \bar{\boldsymbol{v}}\times\boldsymbol{e}_y\right)\times\boldsymbol{e}_y\,, \tag{28}$$

where

$$\bar{\boldsymbol{J}}\times\mathbf{e}_y = \left(-\nabla\bar{\phi} + \bar{\mathbf{v}}\times\mathbf{e}_y\right)\times\mathbf{e}_y = -\frac{\partial\bar{\phi}}{\partial z}\mathbf{e}_x + \frac{\partial\bar{\phi}}{\partial x}\mathbf{e}_z - \bar{v}_x\mathbf{e}_x - \bar{v}_z\mathbf{e}_z\ . \tag{29}$$

Acting by the operator $rot$ on the last term of Eq. (1) and averaging it, we obtain

$$rot\,\overline{\left[\boldsymbol{J}\times\mathbf{e}_y\right]} = \overline{(\mathbf{e}_y\cdot\nabla)\boldsymbol{J}} = J_y(y=-W/2) - J_y(y=W/2)\ . \tag{30}$$

Therefore, the solenoidal part of the last terms of the averaged momentum equation (27) and (28) are defined by the electric current at the spanwise (Hartmann) boundaries. Apparently, the solenoidal part is the only one affecting the velocity, while the potential part contributes only to the pressure. In the case of electrically insulating spanwise boundaries, i.e., the boundary conditions (11b), the right hand side of Eq. (12) is zero, and the last term of Eq. (28) is potential. Introducing its scalar potential as

$$\left(-\overline{\nabla\phi} + \bar{\boldsymbol{v}}\times\boldsymbol{e}_y\right)\times\boldsymbol{e}_y = -\nabla\varphi\,, \tag{31}$$

the momentum equation (28) becomes

$$\frac{\partial \bar{\boldsymbol{v}}}{\partial t} + (\bar{\boldsymbol{v}}\cdot\nabla_\perp)\bar{\boldsymbol{v}} = -\nabla_\perp\left(\bar{p} + \frac{Ha^2}{Gr}\varphi\right) + \frac{1}{Gr^{1/2}}\Delta_\perp \bar{\boldsymbol{v}} - \frac{Hd}{Gr^{1/2}}\bar{\boldsymbol{v}} + \bar{T}\boldsymbol{e}_z\,, \tag{32}$$

and we arrive to the well-known Q2D model considered in [1-7,1922-23].

Assuming the boundary conditions (12), the authors of [22] evaluated the last term of Eq. (28) and arrived to the averaged model having the damping term proportional to $Ha^2$. Further comparison of results yielded by their averaged model and the full three-dimensional problem showed a considerable disagreement.

To gain a Q2D averaging of the last term of the momentum equation (28) for perfectly conducting spanwise boundaries, we recall the equation defining the electric potential in the fully 3D model. Applying the operator $div$ to the Eq. (2), and using Eq. (4), we obtain

$$\Delta\phi = \frac{\partial^2\phi}{\partial x^2} + \frac{\partial^2\phi}{\partial y^2} + \frac{\partial^2\phi}{\partial z^2} = \frac{\partial v_x}{\partial z} - \frac{\partial v_z}{\partial x} . \tag{33}$$

In the case of perfectly conducting boundaries the potential must decay at the spanwise boundaries $y = \pm W/2$. The simplest assumption would be that its $y$-dependence is similar to that of the velocity, i.e., $\phi(x,y,z) = \bar{\phi}(x,z)h(y)$. Then Eq. (33) and its $y$-average become (note that $\overline{h''(y)} \xrightarrow[Ha\to\infty]{} -2Ha/W = -Hd$)

$$\Delta\phi = \left(\frac{\partial^2\bar{\phi}}{\partial x^2} + \frac{\partial^2\bar{\phi}}{\partial z^2}\right) h(y) + \bar{\phi}(x,z)h''(y) = \frac{\partial v_x}{\partial z} - \frac{\partial v_z}{\partial x} = \left(\frac{\partial \bar{v}_x}{\partial z} - \frac{\partial \bar{v}_z}{\partial x}\right) h(y) , \tag{34}$$

and after averaging and for $Ha \to \infty$

$$\left(\frac{\partial^2\bar{\phi}}{\partial x^2} + \frac{\partial^2\bar{\phi}}{\partial z^2}\right) - Hd\bar{\phi}(x,y) = \left(\frac{\partial \bar{v}_x}{\partial z} - \frac{\partial \bar{v}_z}{\partial x}\right) . \tag{35}$$

After $\bar{\phi}(x,z)$ is obtained, the averaging of the gradient of potential $\phi$ yields

$$Ha^2\overline{\nabla\phi} = Ha^2\overline{h(y)}\nabla_\perp\bar{\phi}(x,z) + Ha^2\bar{\phi}(x,z)\overline{h'(y)}\mathbf{e}_y = Ha^2\nabla_\perp\bar{\phi}(x,z), \tag{36}$$

and

$$Ha^2\overline{\nabla\phi} \times \mathbf{e}_y = Ha^2\left(\frac{\partial\bar{\phi}}{\partial z}\mathbf{e}_x - \frac{\partial\bar{\phi}}{\partial x}\mathbf{e}_z\right) . \tag{37}$$

The $y$-averaged momentum equation for the case of perfectly electrically conducting boundaries reads

$$\frac{\partial \bar{v}}{\partial t} + (\bar{\boldsymbol{v}}\cdot\nabla_\perp)\bar{\boldsymbol{v}} = -\nabla_\perp\bar{p} + \frac{1}{Gr^{1/2}}\Delta_\perp\bar{\boldsymbol{v}} + \bar{T}\boldsymbol{e}_z - \frac{Hd}{Gr^{1/2}}\bar{\boldsymbol{v}} - \frac{Ha^2}{Gr}\left(\frac{\partial\bar{\phi}}{\partial z}\mathbf{e}_x - \frac{\partial\bar{\phi}}{\partial x}\mathbf{e}_z + \bar{\boldsymbol{v}}\right), \tag{38}$$

and must be solved together with Eq. (35). The momentum equation for the conducting boundaries contains an additional damping term proportional to $Ha^2$, similarly to the findings of [22]. Note that in the limit $W \to \infty$, and a constant $Ha$, the parameter $Hd$ vanish. Therefore, in the case of insulating spanwise boundaries, the electromagnetic damping will vanish as well, however in other cases the damping will be govern by the last term of Eq. (38) proportional to $Ha^2$.

It should be noted here that if electric conductivity of the spanwise boundaries is finite, the electric potential does not vanish at the borders, and assumption that its $y$-dependence is described by the profile $h(y)$ is wrong. The same can be said about temperature, since if the dimensionless temperature usually does not vanish at the spanwise boundaries, so that its $y$-dependence also is not described by the profile $h(y)$. Derivations in [1-7,21-23] did not address this issue. To average the convective term in the energy equation, we again neglect the term containing $v_y$, and use the representation (19). This yields

$$\overline{(\boldsymbol{v}\cdot\nabla)T} \approx \overline{(h(y)\overline{\boldsymbol{v}}\cdot\nabla)T} = \overline{([\overline{\boldsymbol{v}}-(1-h(y))\overline{\boldsymbol{v}}]\cdot\nabla_{\perp})T} = (\overline{\boldsymbol{v}}\cdot\nabla_{\perp})\overline{T} - \overline{\big((1-h(y))\overline{\boldsymbol{v}}\cdot\nabla_{\perp}\big)T} \quad (39)$$

Noticing that $h(y) \underset{Ha\to\infty}{\longrightarrow} 1$, we neglect the last term of the above equation. For the averaging of the other terms, we use the boundary conditions (10), and obtain

$$\frac{\partial\overline{T}}{\partial t} + (\overline{\boldsymbol{v}}\cdot\nabla_{\perp})\overline{T} = \frac{1}{PrGr^{1/2}}\Delta_{\perp}\overline{T}. \quad (40)$$

Note, that for derivation of the averaged energy equation (40), no assumptions about $y$-dependence of the temperature field were needed. At the same time, if the spanwise boundaries are not perfectly thermally insulated, averaging of the conducting term of the energy equation (10) will contain an additional term, i.e., the difference of normal derivatives of the temperature at the spanwise borders.

## 3. Numerical approach

Both fully 3D and Q2D problems were discretized using the same finite volume schemes as those applied in [23,28-30]. The Q2D steady states were calculated using the Newton method, as described in [2829]. The steady 3D flows were calculated by the time integration using the semi-implicit three time-layers fractional time step scheme with the TPF and TPT direct solvers for the Helmholtz and Poisson equations [30,31].

Since at large Hartmann numbers thin boundary layers develop along all the boundaries, the grid stretching near the boundaries is mandatory. It was found that the stretching needed for correct representation of the 3D flows must be extremely steep [32]. The stretching functions used in [32], e.g.,

$$x \to 0.5 + 0.5\frac{tanh[s(x-0.5)]}{tanh(0.5s)} \quad (41)$$

where $s$ is the stretching parameter, were applied for all three coordinates. The stretching transformation (42) allows for the denser grid nodes at the borders $x = 0$ and $x = 1$ for the interval $0 \le x \le 1$. With the increase of the stretching parameter $s$, more grid points are clustered near the boundaries, so that distances between neighbor grid points decrease near the boundaries and increase in the central part of the flow region.

Following [28,32], we monitor the total kinetic energy of the flows obtained as a solution of the full 3D model, and the Q2D model

$$E_{kin,3D} = \int_0^A \int_0^W \int_0^1 \mathbf{v}^2 dxdydz, \qquad E_{kin,Q2D} = \int_0^A \int_0^1 \mathbf{v}^2 dxdz, \tag{42}$$

and the Nusselt numbers on the hot boundaries of the 3D and Q2D configurations

$$Nu_{3D} = \frac{1}{AW} \int_0^A \int_0^W \left(\frac{\partial T}{\partial x}\right)_{x=0} dydz, \quad Nu_{Q2D} = \frac{1}{A} \int_0^A \left(\frac{\partial T_{Q2D}}{\partial x}\right)_{x=0} dz \ . \tag{43}$$

The grid convergence for different stretching parameters was studied in [32] for the test case of convection in the cubic cavity, $A = W = 1$, $Pr = 0.054$, $Ra = GrPr = 10^6$, taken from [26]. Flows under action of the magnetic field with $Ha = 100$ and three different field directions were examined. Following these results, in the present work we adjust the stretching parameter for each certain case, so that in different calculations it is varied between 6 and 15. The stretching parameter used is reported in figure captions below.

The Q2D model does not require so steep stretching because the boundary layers thicken after the $y$-averaging procedure. The computations for the Q2D model in the square cavity were carried out on $200^2$ grid and $s = 4$ in the $x$- and $z$- directions. In the case of horizontally elongated cavities, the number of points in the long direction was increased by the value of aspect ratio. For the 3D problems the maximal number of grid points was $AN_z \times WN_z \times N_z$, where number of points in the z-direction varied between 100 and 500. The grid size was then chosen such that the values reported below were converged to within 2 decimal places at least.

## 4. Results

Two compare fully 3D solutions with the Q2D results, we consider two test configurations. The first one is taken from [26], and was treated as a test problem in [32]. Convection in a laterally heated box with a square cross-section, with the Prandtl number $Pr = 0.054$ and the Rayleigh number $Ra = GrPr = 10^6$ is considered. The second problem is taken from [22]. It is also convection in a laterally heated box, but with a smaller Prandtl number $Pr = 0.015$, the cross-section aspect ratio (length/height) $A = 8$, and the Grashof number $Gr = 10^7$. The width ratio is taken as $W = 1$ for all the three problems in the first series of numerical experiments. In the second series of experiments, the width ratio is varied together with the Hartmann number, so that the dimensionless parameter $Hd = 2Ha/W$ is kept constant.

*4.1. Comparison of fully 3D flows with Q2D model ones*

The total kinetic energy and the Nusselt numbers defined in Eqs. (42) and (43) and calculated using the full 3D and Q2D models are compared in Table 1 for the width ratio $W = 1$ and two aspect ratios $A = 1$ and 8. Three cases of different boundary conditions for the electric potential were all the boundaries are either perfectly insulated or perfectly conducting, and the case when the spanwise boundaries are perfectly insulated while the others are perfectly conducted. It follows from Table 1 that results of the 3D and Q2D models compare very well when the spanwise boundaries are electrically insulated, independently on the electrical boundary conditions on the other four boundaries. In the case of electrically conducting spanwise boundaries, results of the two models are noticeably different. The only common feature observed, is a strong flow damping at large Hartmann numbers, at which the kinetic energy tends to zero and the Nusselt numbers tends to their values corresponding to the purely thermally conducting state. Note also, that the Nusselt numbers calculated for the configuration of [26] with $Pr = 0.054$, using the fully 3D and Q2D models agree much better than those reported in [1924]. Most probably, this is because of a better 3D spatial resolution and steep grid stretching applied here.

To illustrate the flow patterns, we plot three-dimensional isosurfaces of velocities and isotherms of flows for the parameters of [26] at $Ha = 100$, for either electrically insulating or electrically conducting boundaries (Figs. 3 and 4). The temperature and velocity patterns in the case of the spanwise electrically insulated boundaries and the other boundaries electrically conducting, are similar to those plotted in Fig. 3 and are not shown. First of all, we observe that in both cases the $y$-component of velocity is noticeably weaker than the two others, and is antisymmetric with respect to the midplane $y = 0.5$, so that its spanwise average (13) is zero (frames (b) in Figs. 3 and 4). Second, we observe that the horizontal and vertical velocity components, as well as the temperature, almost do not change far from the borders $y = 0$ and $y = 1$ (frames (a) and (c) of Figs. 3 and 4). A large change, however, is observed in the isothermal surfaces, which, in the case of electrically insulating borders, form more profound boundary layers near the hot and cold boundaries (cf. Fig. 3d and Fig. 4d). However, since the velocity patterns are similar, all these observations do not explain why the Q2D model becomes problematic when the spanwise boundaries are electrically conducting.

To gain a more detail description of the velocity distribution, we plot one-dimensional velocity profiles in the midplane $x = 0.5$, and different heights (Fig. 5). In both cases of the electrically insulating spanwise boundaries (Fig. 5a-5d), the velocity profiles are not strictly monotonic, but nevertheless are quite close to the idealistic assumption of the profile $h(y)$ shown in Fig. 2. The latter is not true for the case of all conducting boundaries (Fig. 5e and 5f), where velocity profiles exhibit a noticeable change far from the borders $y = \pm 0.5$. Apparently, the assumption the profile $h(y)$ in Eq. (19) should be amended for this case. The results below support this conclusion.

To strengthen the above arguments, we plot equipotential surfaces (Fig. 6) and profiles of the electric potential (Fig. 7) for the three cases considered. In the cases of all or only spanwise boundaries electrically insulated, the potential is distributed similarly (cf. Fig. 6a and 6b). In the case of all boundaries insulated, its profiles resemble the profile $h(y)$ (Fig. 7a), but this does not happen in the two other cases (Figs. 7b and 7c). As is explained above, in the case of electrically insulated spanwise boundaries, the electric potential affects only pressure and does not affect velocities, so that its distribution does not affect the validity of Q2D model. In the case of electrically conducting spanwise boundaries, the potential affects the velocity as well. Figure 7c shows that the assumption of its $y-$dependence made in Eq. (32) is wrong.

The remaining question is how the above conclusions change if the magnetic field is strongly increased, say to $Ha = 1000$. The results reported in Table 1 show that the above conclusion holds. Namely, the results of Q2D model remain closed to the fully 3D results when the spanwise boundaries are electrically insulated. Figures 8 and 9 show isosurface of the velocity components and temperature for the cases of insulating and conducting boundaries. In both cases the $u$ and $w$ velocity components are considerably larger than $v$ component. They are almost constants in the bulk of the boundary layers and swiftly decay towards the spanwise boundaries. Isosurfaces corresponding to small levels of velocity components illustrate additionally that in the case of insulating boundaries (Fig. 8) the velocity are almost independent on $y$, while in the conducting boundaries case (Fig. 9) a certain dependence is observed in the bulk of the flow.

Another important observation from Figs 8 and 9 is much smaller velocity values in the conducting boundaries case, in which, as already was stated above, the flow is strongly damped. The latter is seen also in the temperature isosurfaces, which are almost vertical planes in Fig. 9d, meaning that convective heat transfer is negligible and the liquid approaches the purely heat conduction state. Contrarily, when the spanwise borders are electrically insulated, the temperature isosurfaces are curved and exhibit shapes characteristic for this kind of convection flow (see e.g., [28]).

The difference between the insulating and conducting boundaries is seen better in the one-dimensional velocity profiles shown in Fig. 10. In the case of the insulating spanwise boundaries (Figs. 10a-d), the velocities are almost $y$-independent in the largest part of the flow region except very thin boundary layers, where they steeply decay towards the borders. If the borders are electrically conducting, the results differ, and velocity profiles far from the boundaries still do not tend to the horizontal lines (Fig. 10e,f).

A similar observation takes place for profiles of the electric potential (Fig. 11). In the case of conducting spanwise boundaries (Fig. 11a,b), the potential is almost a constant along the $y$-coordinate, so that one can assume it to be a function of $x$ and $z$ only. When spanwise boundaries are conducting, (Fig. 11c), there is an obvious dependence on the spanwise coordinate, and the corresponding potential profiles are quite different from what is assumed by the profile $h(y)$ (see Fig. 2 and Eq. (19)).

The conclusions remain similar for a horizontally elongated cavity with $A = 8$. Comparing the total kinetic energy for the fully 3D solution and Q2D model (Table 1), we observe again that the results are close when the spanwise boundaries are electrically insulated. In the case of electrically conducting boundaries the results noticeably differ, with an exception for $Ha = 1000$, when the flow is so strongly damped that the kinetic energy becomes extremely small, of the order of $10^{-6}$, while the Nusselt number approaches its thermal conduction value, which is 0.125. One can attribute this strong damping to the term proportional to $Ha^2$ in Eq. (36), but it should be stressed that the agreement is only qualitative, but not quantitative.

Spatial distributions of the velocity and the temperature in the elongated cavity (Figs. 12-14) are similar to what was observed for the cubic one. The flow at $Ha = 1000$ and conducting spanwise boundary is almost fully suppressed and is not shown. In all the three figures, we observe again that the spanwise velocity is much weaker than the horizontal and vertical ones. It is antisymmetric with respect to the midplane $y = 0$, which makes its average by eq. (13) zero. We observe also thin boundary layers near the horizontal and vertical boundaries, forming by, respectively, the horizontal and vertical velocities, which develop in cavities with either insulating or conducting spanwise boundaries. These velocities are almost constants far from the spanwise boundaries, which justifies use of the profile (19) for the Q2D model. As follows from Table 1, the values of the Nusselt number and kinetic energy are close to those predicted but the Q2D model. It should be noted, however, that in the case of electrically conducting spanwise boundaries, the agreement is due to the strong electromagnetic damping, and not because of validity of the Q2D formulation. It should be noted also, that the flows consist of only one convective circulation, so that a multi-cellular flow patterns, which can be expected in an elongated cavity [33], are suppressed.

An example of the equipotential surfaces in the elongated cavity is shown in Fig. 15. Similarly to the cubic cavity (Fig. 6), the isosurfaces are almost identical for the two cases with electrically insulated spanwise boundaries. In the case of electrically conducting spanwise boundaries, we observe that electric potential is two orders of magnitude weaker. Assuming that that the terms with electric potential in Eq. (38) can be neglected, the last term $-Ha^2\boldsymbol{v}$ describes the damping force directed against velocity everywhere. Since at large Hartmann numbers, $Ha^2 \gg Hd = 2Ha/W$, this term dominates and yields a very strong damping observed in the case of electrically conducting spanwise boundaries.

Finally, we calculate the *y*-averaged flows from the 3D results and compare them with the results of Q2D model. It is important to notice that if the numerical scheme resembles the problem symmetries, the averaging will result in a two-dimensional vector $\bar{\mathbf{v}}^{3D} = (\bar{v}_x^{3D}, 0, \bar{v}_z^{3D})$. Also, with the use of an appropriate quadrature formula for the averaging, the discretized (numerical) divergence of $\bar{\mathbf{v}}^{3D}$ vanishes, similarly to Eq. (14). The latter allows one to represent the vector $\bar{\mathbf{v}}^{3D}$ by a stream function $\bar{\psi}^{3D}$, which can be compared with the stream function of the Q2D solution. The comparisons are shown in Fig. 16 for the cubic, and in Fig. 17 for the horizontally elongated cavity, all with insulating spanwise boundaries.

In the case of the cubic cavity (Fig. 16), the streamlines are close and the isotherms are identical for $Ha = 100$ (Fig. 16a), which is illustrated for $W = 1$ and $W = 5$. At $Ha = 1000$, the streamlines and isotherms coincide for $W = 1$, as well as for $W = 10$ (Fig. 16b). It is emphasized that to obtain these close results we had to apply a very aggressive stretching, $s = 10$ for $Ha = 100$, and even more aggressive $s = 15$ for $W = 5$. Note that the maximal values of stream functions are reported in caption of Fig. 16 and are also close.

The above comparison made for the horizontally elongated cavity is shown in Fig. 17. We observe that at larger magnetic field $Ha = 1000$ (Fig. 17b), the result agree, however to obtain this agreement we had to apply a very aggressive stretching $s = 16$. At lower Hartmann number, $Ha = 100$ (Fig. 17a), the maximal values of the stream function differ for Q2D model (black lines) coincide with the results for fully 3D flow with electrically insulated spanwise boundaries (red and blue lines). For the cavity with a large width ratio $W = 10$, the streamlines and isotherms are close, but maximal value of the stream function is smaller. As will be shown below, this happens because the stretching parameter $s = 10$ used for this case is insufficient for large width ratios. With a steeper stretching, $s \geq 13$, the time-dependent calculations result in a chaotically oscillating flow, which does not allow us to make an accurate time averaging and to extract the two-dimensional spanwise averaged stream function.

An additional comparison of fully 3D and Q2D model is presented in Fig. 18 for $W = 1$, which replicates Figs. 3 and 6 of [22]. In the case of electrically insulated boundaries (Fig. 18a), the profiles are very close. In the case of electrically conducting boundaries (Fig.18b) the profiles are close far from the top and the bottom. Near the horizontal boundaries $z = 0$ and $z = 1$, the Q2D model overestimates the electromagnetic damping, and predicts significantly lower velocities compared to those obtained in the fully 3D computation.

### *4.2. Asymptotics at large width ratios*

Since the Q2D model is independent on the box width ratio $W$, and is validated above mainly for $W = 1$, it will be applicable to larger aspect ratios if there exists an asymptotics for $W \to \infty$, such that the final result is not far from that at $W = 1$. To keep the Q2D result unchanged, the parameter $Hd = 2Ha/W$ must be kept constant when the width ratio $W$ varies, meaning that the Hartmann number grows proportionally to the width ratio. This asymptotic behavior can be expected for electrically insulated spanwise boundaries, for which the Q2D averaged equation (32) is valid. Nevertheless, it would be worth to examine whether a similar asymptotic behavior can be observed also for electrically conducting spanwise boundaries. This can happen, for example, if at large $W$ the term proportional to $Ha^2$ in Eq. (28) becomes small compared to the term proportional to $Hd$. The latter can be expected, for example, if the effect of spanwise boundary conditions diminishes at large width ratios, so that the difference between the electrically conducting or insulating boundaries becomes negligible, and the term in brackets proportional to $Ha^2$ in Eq. (28) tends to zero.

To verify existence of the asymptotic limit of the spanwise averaged flow, we monitor several scalar characteristic values derived from the flow and temperature fields. The first such characteristic is the Nusselt number $Nu_{3D}$. The second one is the total kinetic energy per unit width, which is given by $E_{kin}/W$. The third one is the Nusselt number at the midline $x = y = 0$, defined as

$$Nu_{midline} = \frac{1}{A}\int_0^A \left(\frac{\partial T}{\partial x}\right)_{x=0,y=0} dz, \qquad (44)$$

which can be compared with the Nusselt number of the two-dimensional Q2D model calculated at the boundary $x = 0$. The fourth one is the maximal value of the stream function of the $y$- averaged flow $\bar{\psi}^{3D}$. Note, that we can also define the Nusselt number of the spanwise averaged temperature field as

$$\overline{Nu} = \frac{1}{A}\int_0^A \left(\frac{\partial \bar{T}}{\partial x}\right)_{x=0} dz \ . \tag{45}$$

Assuming that dependence of the temperature on the *y*-coordinate also can be described as $T(x, y, x) \approx h_T(y)\bar{T}(x, z)$, with $h_T(y) \neq h(y)$ and $\overline{h_T(y)} = 1$, it can be easily seen that $\overline{Nu}_{3D} = \overline{Nu}$. We observe this equality in the calculations for both insulating and conducting boundaries.

As noted above, the stream functions of the Q2D model and of the spanwise average 3D velocity field can be directly compared, so that the fourth characteristic $\psi_{max,average} = max|\bar{\psi}^{3D}|$ can be derived by the spanwise averaging of the 3D numerical result. Thus, all the four characteristic can be compared with their Q2D counterparts, which yields another verification of the Q2D model. The four above characteristics were monitored for different sets of parameters. The results are presented in Figs. 19-21.

We start from the cavity of square cross-section, $A = 1$ and $Hd = 200$ (Fig. 19a,b). When the spanwise boundaries are electrically insulated (Fig. 19a), with increase of the width ratio $W$, the three-dimensional characteristics approach values close to their Q2D counterparts that are shown inside the frames. At the same time, there appears a problem, since beyond $W = 2$ the flow becomes oscillatory. The corresponding values, shown in the graphs, are calculated by, first, time averaging over the oscillation period, and then *y*-averaging along the spanwise coordinate. Thus, there arises another question: whether the Q2D model predicts not only time-averaged flow properties, but also its oscillatory instability. This question is addressed in the next section. Meanwhile, we observe that in spite the averaged characteristic do not exhibit a clearly asymptotic behavior, their values remain very close to those predicted by the Q2D model.

When the spanwise boundaries are electrically conducting (Fig. 19b), with increasing $W$, all four characteristics approach asymptotic values, but they differ from those predicted by the Q2D model. This example shows that in the case of conducting spanwise boundaries the above Q2D model needs amendments, possibly the spanwise profile (19) should be amended, or spanwise profiles of velocity and electric potential should be different (see Fig. 11c).

Figure 19c shows results for the insulating spanwise boundaries and larger magnetic field, with $Hd = 2000$. In this case we observe that at very strong stretching, $s = 15$, the characteristic values only slightly vary with variation of the width ratio, and remain close to those predicted by the Q2D model. When smaller stretching parameters are applied, the characteristic values tend to decrease with the growing width ratio, and the decrease is steeper for smaller $s$. This example shows how critical the stretching can be, especially at large Hartmann numbers. Note, that the result for $W = 10$ in Fig. 17b (green lines) was obtained with $s = 16$, while the results calculated with $s = 10$ and 13 were not correct. We can suggest here, that successful comparison with the Q2D model can be an argument for the choice of the appropriate stretching.

In the case of conducting spanwise boundaries and $Hd = 2000$, the flow is almost completely damped, so that kinetic energy is very small, and the Nusselt number is very close to unity. Graphs for this case are not reported.

Figures 20 and 21 show how the four scalar flow characteristics change with variation of the width ratio $W$ in elongated cavities with $A = 8$ for $Hd = 200$ and 2000, respectively. Like in the previous case, configuration with the electrically insulating boundaries (Figs. 20a and 21a) exhibit necessity of a strong stretching. When stretching with $s = 13$ is applied in the case of $Hd = 200$, and $s = 16$ for $Hd = 2000$, the solutions chaotically oscillate in time. Since there is no an oscillation period, an accurate time averaged value cannot be derived. The results of time averaging over a very long time (several hundred dimensionless time units) are shown on the frames (a) by diamonds. Again, we observe that results of these averaging are quite close to the Q2D predictions. To illustrate importance of the strong stretching, the results obtained for $s = 10$ are presented as circles connected by the solid curves. The filled circles correspond to oscillatory states with a definite period, so that their average value can be accurately evaluated. Here again, we observe that insufficient stretching results in decaying of the characteristic values with the increase of the width ratio.

In the case of conducting spanwise boundaries (Figs. 20b and 21b), we observe a clear asymptotics at large aspect ratios. We note, that the values of all four characteristics are noticeably smaller than those of the insulated boundaries case. This means that with the conducting spanwise boundaries, the flow is much stronger damped, so that at $Hd = 2000$ (Fig. 21b) the value of all the three Nusselt numbers becomes 0.125, which is the Nusselt number of the purely conducting state. As mentioned above, since the electric potential in this case becomes very small, the damping is attributed to the term $-Ha^2\boldsymbol{v}$ in Eq. (38), which becomes dominant since at large Hartmann numbers, $Ha^2 \gg Hd$. Thus, in a sufficiently strong magnetic field the flow is fully damped, so that no any additional or simplified flow model is needed. Surprisingly, at smaller Hartmann numbers, $Hd = 200$ (Fig. 21a), when the flow is not completely damped yet, the asymptotic limit of all the four characteristics are close to the Q2D predictions of the above Q2D model, and the agreement is much better than in the case of $A = 1$ (Fig. 19b). This indicates again that the profile $h(y)$ should be amended to $h(x, y)$ to include dependence on another horizontal coordinate.

*4.3. Oscillatory instability of 3D and Q2D flows*

As mentioned above, studying the configuration of Mößner & Müller [26] with $A = 1$, $Pr = 0.054$, and $Ra = 10^6$, we observed that at $Ha = 100$ the flow becomes oscillatory beyond $W \geq 2$. The oscillatory instability of Q2D convective flows was studied in [23], however the question whether this model can describe also instabilities of fully 3D flows remains open. We computed the critical Rayleigh number for the Q2D flow in square cavity and obtained $Ra_{cr} = 1.08 \times 10^6$, which is very close to the fully 3D results for $W \geq 2$. We cannot establish dependence of the critical Rayleigh number on width ratio, but even if it noticeably varies, the neutral curve $Ra_{cr}(W)$ approaches the Q2D value at $W = 2$.

To compare the Q2D and fully 3D results further we compare frequencies of the computed 3D oscillations with the predictions of linear stability theory applied to the Q2D model. The dimensionless frequency predicted by the Q2D model is approximately 0.50, while in the 3D calculations it varies from 0.521 at $W = 2$ to 0.518 at $W = 5$, so that the values are very close.

Finally, we compare the isosurfaces of the 3D temperature oscillations amplitude with the amplitude of the most unstable perturbation of the temperature in the Q2D model. The result is presented in Fig. 22. We observe that in the middle cross-section $y = 0$, as well as quite far from it, the amplitude isolines resemble the Q2D perturbation ones (cf. Figs. 21a,b with Fig. 21c). This takes place both for the width ratios $W = 2$ and $W = 5$. At the same time, the isosurfaces shape shows that the 3D amplitudes vary with the spanwise coordinate $y$ far from the spanwise boundaries stronger than it was observed, e. g., for the steady flow velocity and temperature profiles (Figs. 5a,b and 7a,b). This means that while the Q2D model can correctly predict the instability onset, while in supercritical flow regimes one can expect the fully 3D and Q2D results are expected to diverge from each other.

## 5. Conclusions

In this study we examined at which conditions Q2D model can yield results well compared with those of the fully 3D numerical modelling, when is applied to convection of electrically conducting fluid in a three-dimensional laterally heated rectangular box under action of a horizontal magnetic field orthogonal to the applied temperature gradient or, in other words, to the main convective circulation.

We recalled derivation of the Q2D model and reiterated conditions, at which this model is theoretically applicable. Thus, the basic Q2D model can be applied to the above 3D configuration of natural convection flow in a 3D box if the magnetic field acts in the spanwise direction (Fig. 1), the initial 3D model is reflection symmetric in the spanwise direction, and the spanwise boundaries are perfectly insulated electrically, as well as thermally. It cannot be applied if the magnetic field has non-zero components in two other directions. If other conditions do not hold, the Q2D model should be reformulated accordingly. When the conditions hold and the Q2D model is valid, the flow is governed by the dimensionless parameter $Hd = 2Ha/W$, which replaces the Hartmann number governing the full three-dimensional formulation.

Analyzing patterns of three-dimensional flows under action of the magnetic field with $Ha = 100$ and $1000$, we observe that in the case of electrically insulating spanwise boundaries, the velocity profiles in the spanwise direction become close to constants far from the borders, which does not happen when the spanwise boundaries are electrically conducting (Figs. 5 and 10). Similar conclusion is made for the spanwise profiles of the electric potential (Fig. 7). This can be the main reason for inapplicability of the discussed Q2D model in this case. We show also that in the case of electrically insulating spanwise boundaries, the Q2D model correctly represents spanwise-averaged flow and temperature fields in cavities with the width ratio $W = 1$ the aspect ratios $A = 1$ and 8 (Figs. 16 and 17).

To study whether the flow becomes dependent solely on the parameter $Hd$, we considered two aspect ratios $A = 1$ and 8, and increased the width ratio $W$, keeping the parameter $Hd$ constant. For scalar flow characteristics, the total kinetic energy, the Nusselt number at the hot border, the 2D Nusselt number at the midplane, and the maximal value of the stream function of the spanwise-averaged flow were monitored. In these calculations the Hartmann number grew with the growth of the width ratio, $Ha = W\ Hd/2$. It appears, that all the four characteristics do exhibit an asymptotic behavior at large width ratios. It takes place even when the flows bifurcate to periodic, as well as chaotic, oscillatory states. It is stressed, however, that to arrive to this conclusion at very large Hartmann numbers, we had to applied a very aggressive grid stretching, so that the stretching parameter in Eq. (41) arrived to $s = 16$. Weaker stretching resulted in wrong results that showed decay of all the four characteristics with the width ratio. We argued that such a comparison can serve as a test for correctly chosen stretching.

In a horizontally elongated boxes with $A = 8$ and perfectly electrically conducting spanwise borders, we also observed asymptotics with the increasing $W$ (Figs. 20b and 21b). We attributed this asymptotics to almost complete electromagnetic damping of the flow. However, existence of such asymptotics at smaller $Hd = 200$ at the aspect ratio $A = 1$ (Fig. 19b), when the flow is not completely damped yet, allows us to assume that an amendment of the Q2D model for electrically conducting spanwise boundaries can also be found.

It should be noted also, that all computations in the horizontally elongated boxes resulted in single convective circulation flows. This shows that the spanwise magnetic field suppresses multicellular multiple flow states reported in [33].

Considering two cavities with the width ratio $W = 1$, and the aspect ratios $A = 1$ and 5, we showed additionally that the Q2D model can correctly predict onset of the oscillatory instability of the flow, including critical Grashof number, frequency of oscillations, as well as spatial distribution of the oscillations amplitude. These examples validate our earlier results obtained on the basis of the Q2D model [23].

It is emphasized that the above conclusions are made on the basis of several particular examples, but they show that even if all the above conditions hold, the Q2D model cannot be applied without validation. Current results indicate on its possible extension for perfectly conducting boundary conditions, but this is left for future studies.

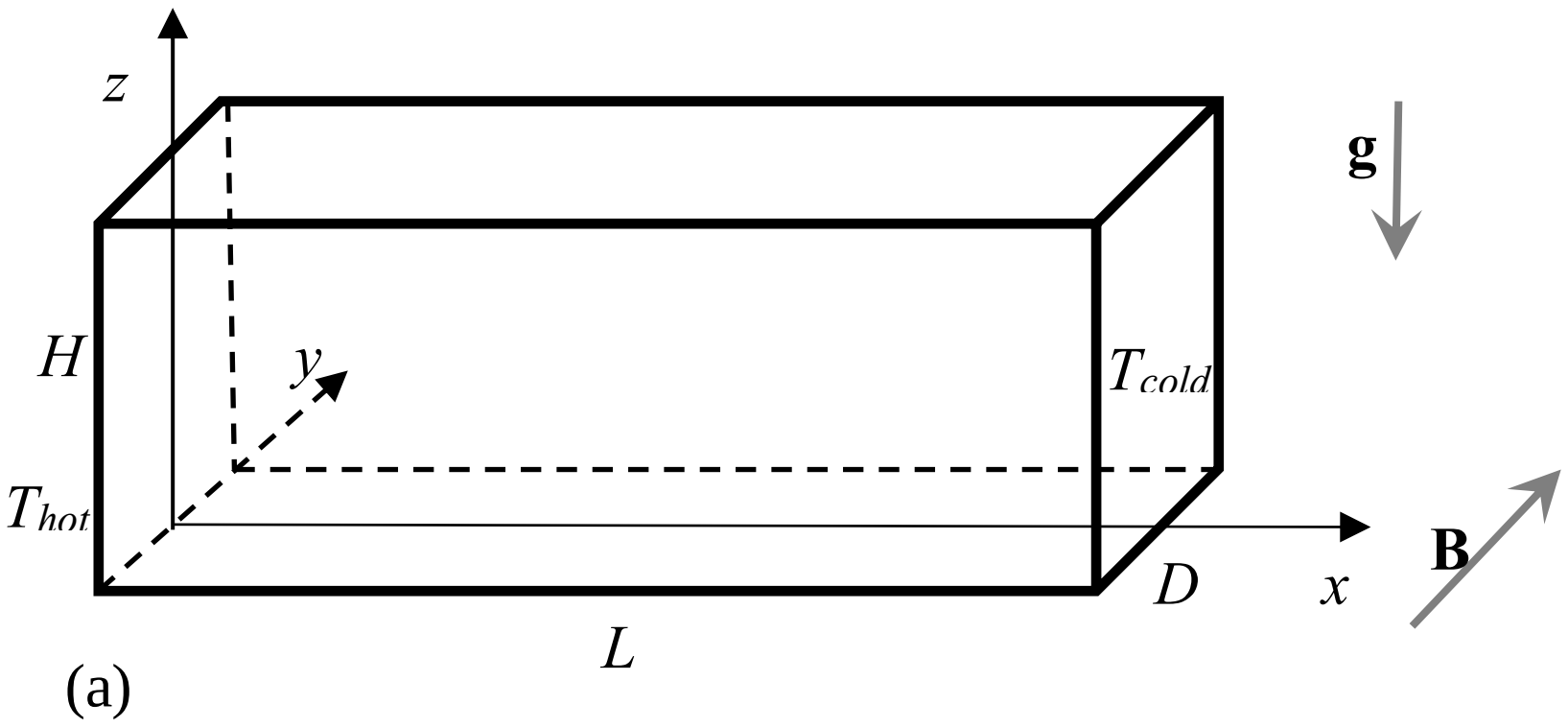

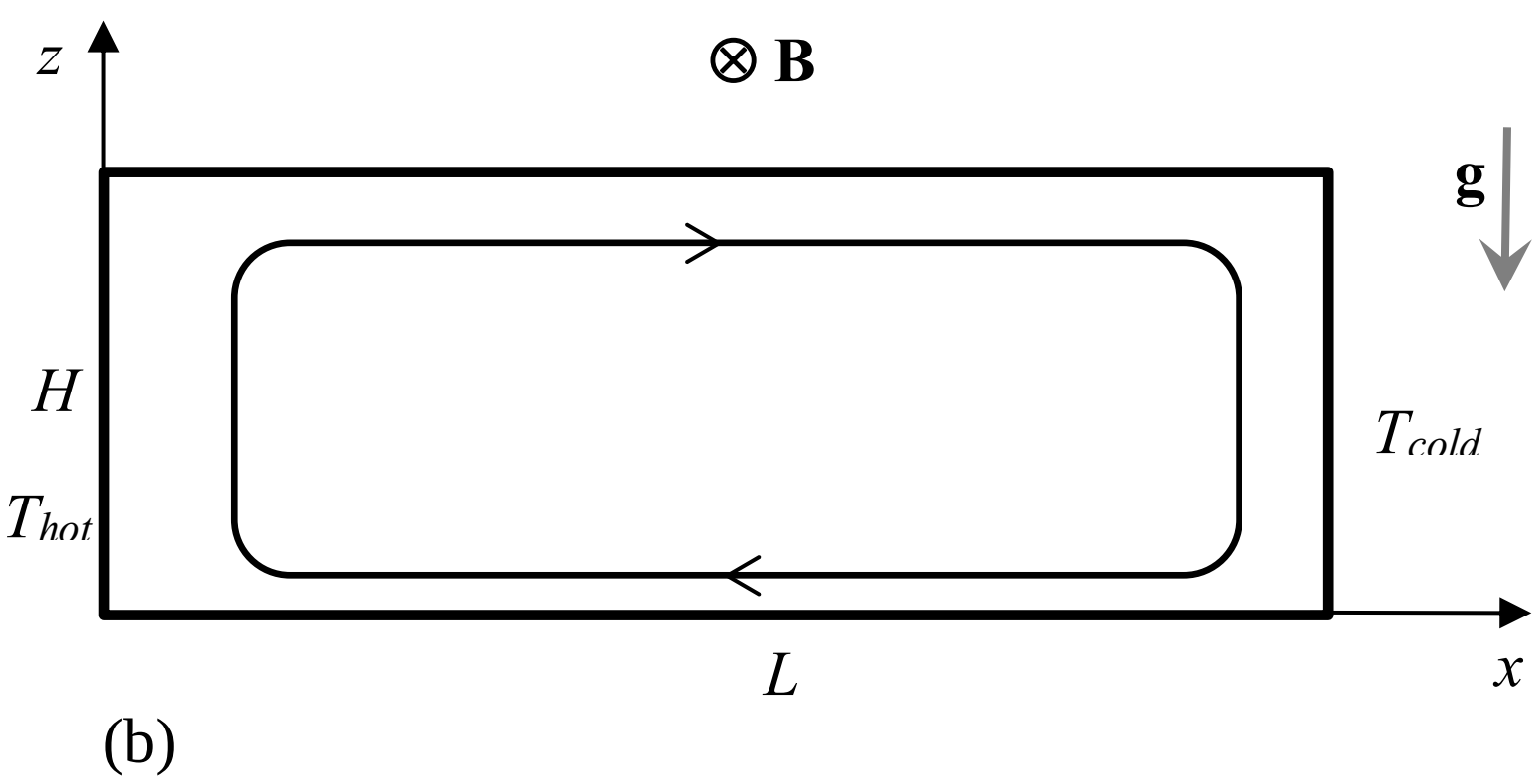

**Figure 1.** Sketch of the considered flow configurations. (a) Laterally heated three-dimensional box under effect of the transverse magnetic field. (b) A laterally heated cavity corresponding to the Q2D two-dimensional model.

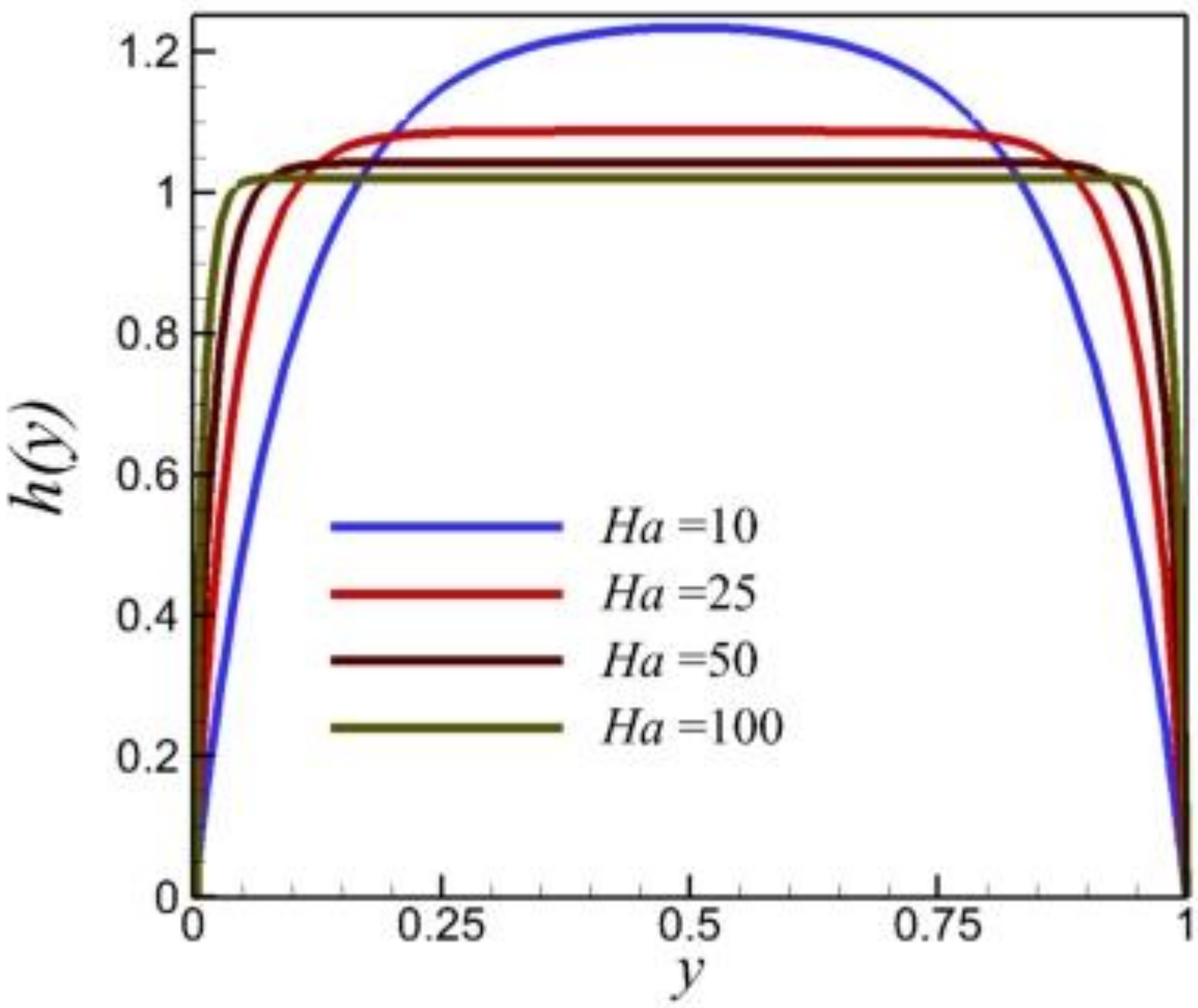

**Figure 2.** Q2D profiles $h(y)$ for different Hartmann numbers.

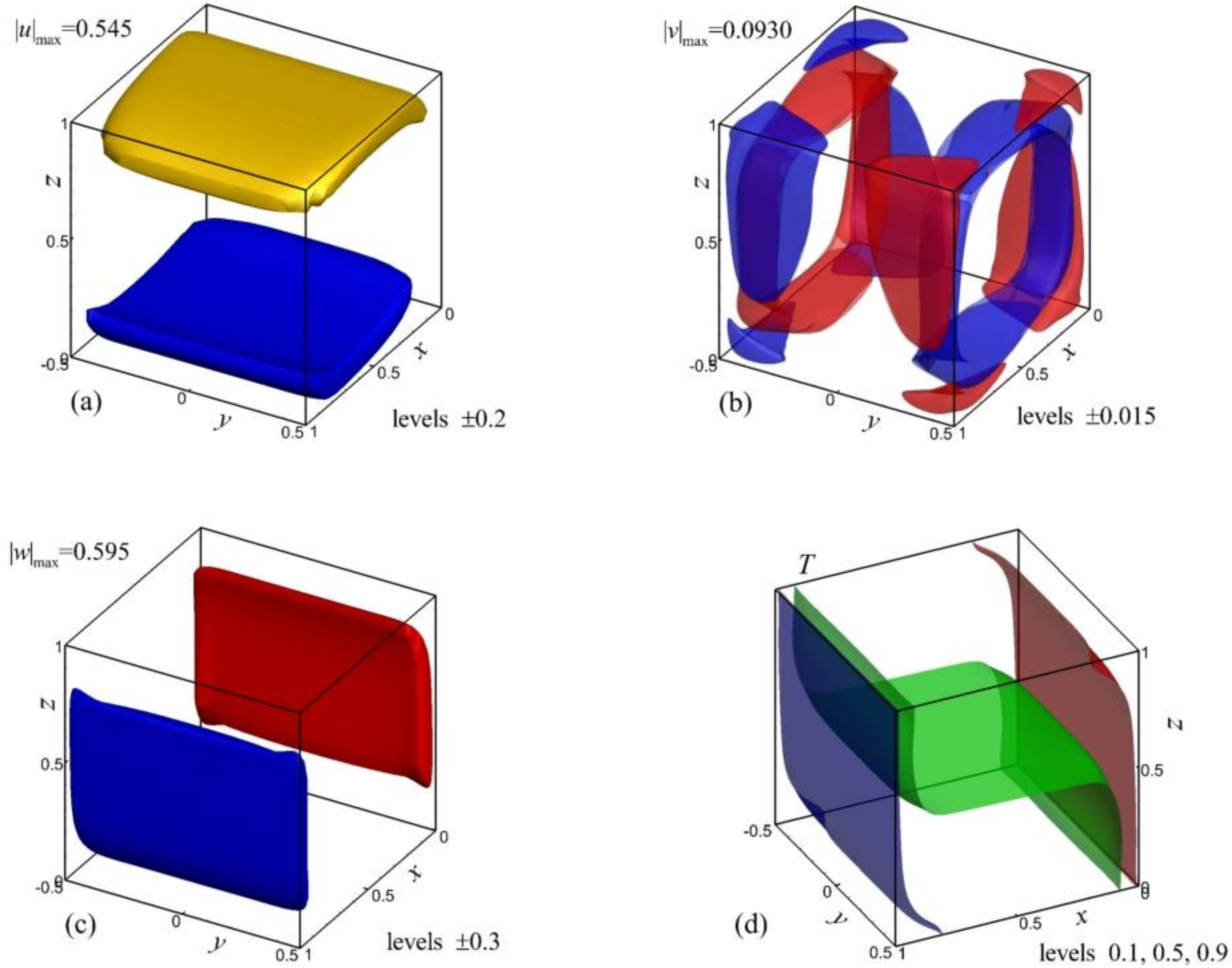

**Figure 3.** Isosurfaces of three velocity components and temperature. Cubic cavity with electrically insulated boundaries.

$A = W = 1, Pr = 0.054,\ \ Ra = GrPr = 10^6, Ha = 100, s = 10.$

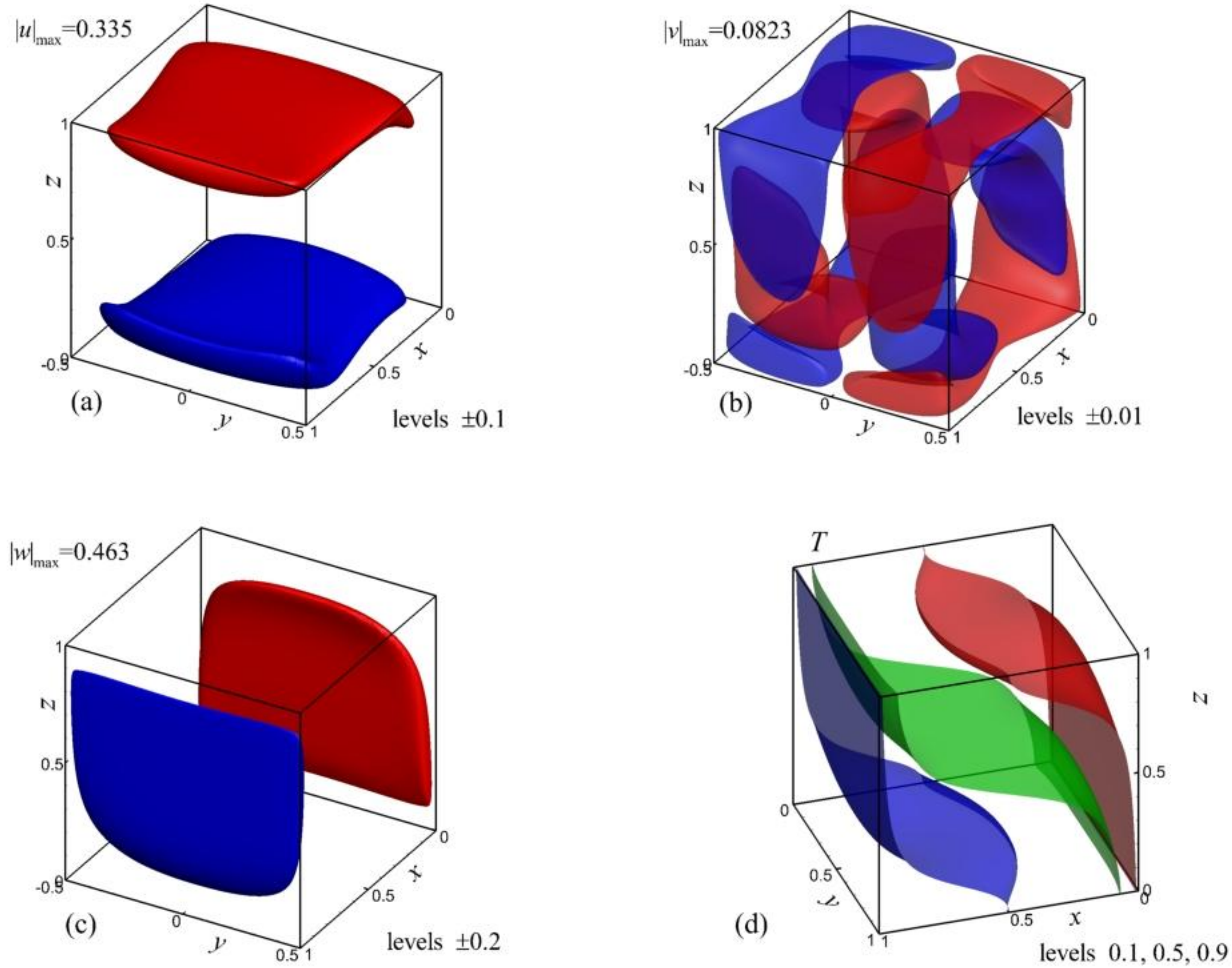

**Figure 4.** Isosurfaces of three velocity components and temperature. Cubic cavity with electrically conducting boundaries.

$A = W = 1, Pr = 0.054, \quad Ra = GrPr = 10^6, Ha = 100, s = 10.$

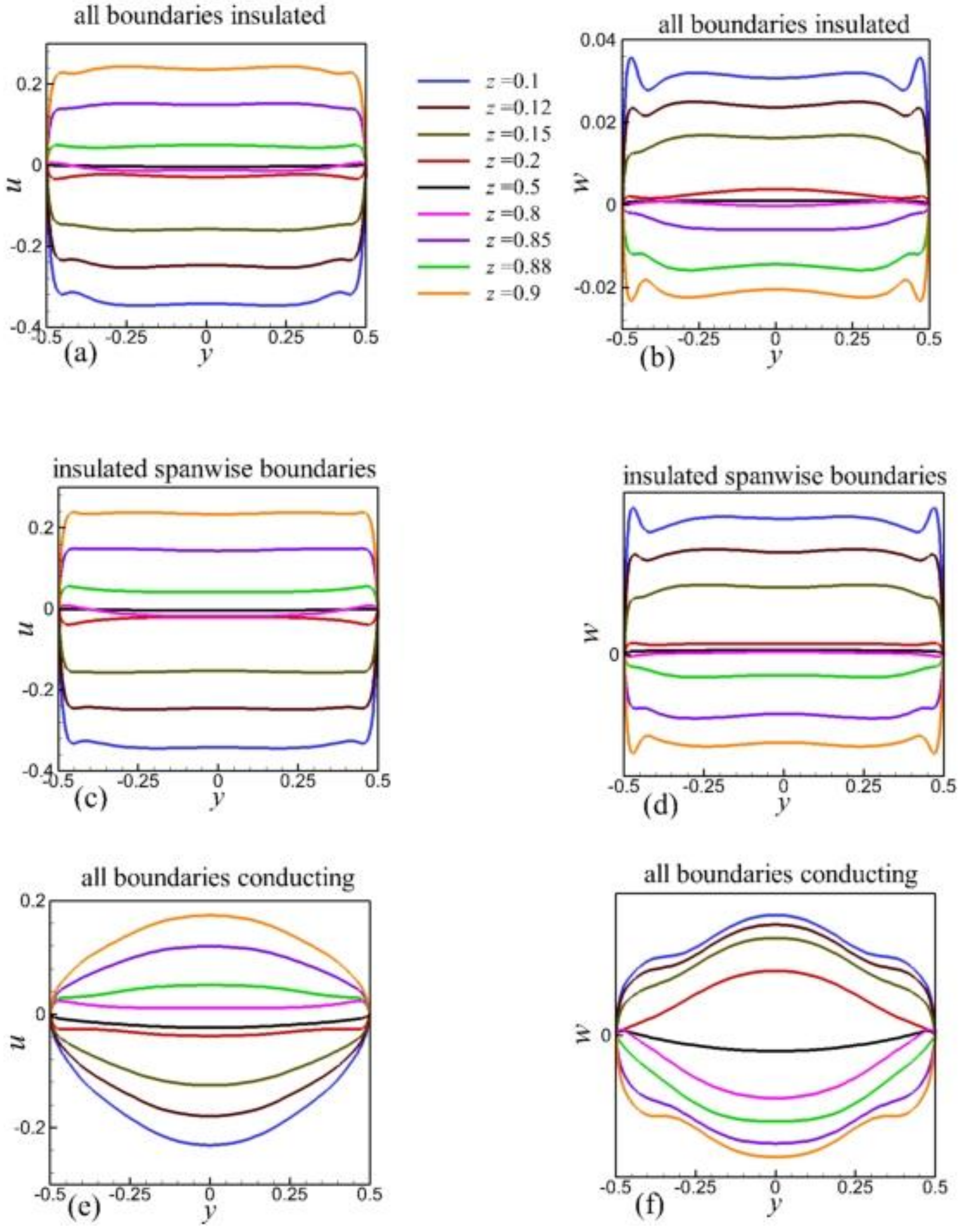

**Figure 5.** Profiles of the horizontal and vertical velocity components along the spanwise coordinate and $x = 0.5$ for three different boundary conditions for the electric potential. (a), (b) – all boundaries are electrically insulated; (c), (d) – spanwise boundaries are electrically insulated, others are electrically conducting; (e), (f) – all boundaries are electrically conducting. $A = W = 1, Pr = 0.054, \ \ Ra = GrPr = 10^6, Ha = 100, s = 10.$

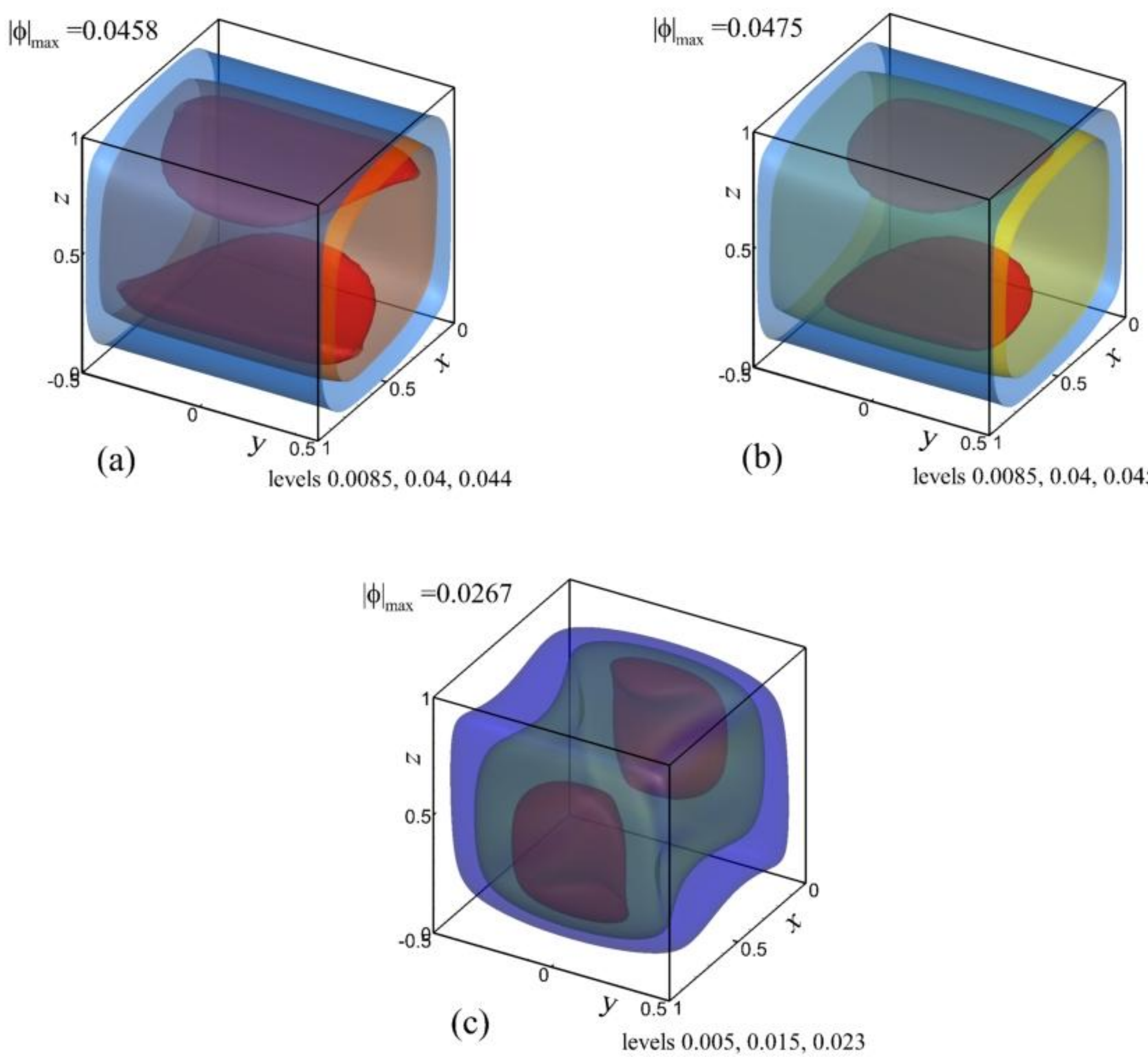

**Figure 6.** Equipotential surfaces for three different boundary conditions for the electric potential. (a) – all boundaries are electrically insulated; (b) – spanwise boundaries are electrically insulated, others are electrically conducting; (c) – all boundaries are electrically conducting. $A = W = 1, Pr = 0.054, \; Ra = GrPr = 10^6, Ha = 100, s = 10$.

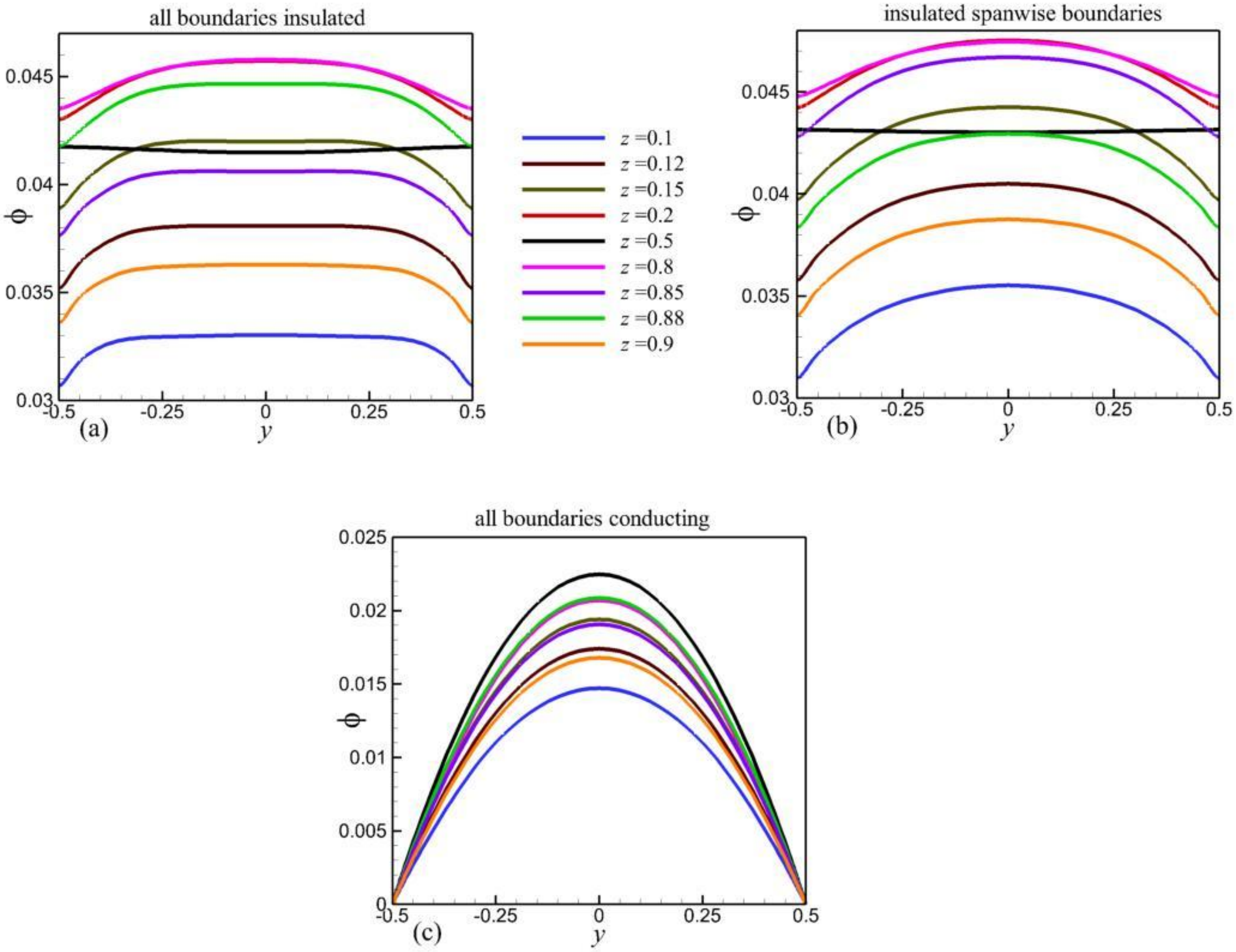

**Figure 7.** Potential profiles along the spanwise coordinate at $x = 0.5$ for three different boundary conditions for the electric potential. a) – all boundaries are electrically insulated; (b) – spanwise boundaries are electrically insulated, others are electrically conducting; (c) – all boundaries are electrically conducting. $A = W = 1, Pr = 0.054,\ \ Ra = GrPr = 10^6, Ha = 100, s = 10.$

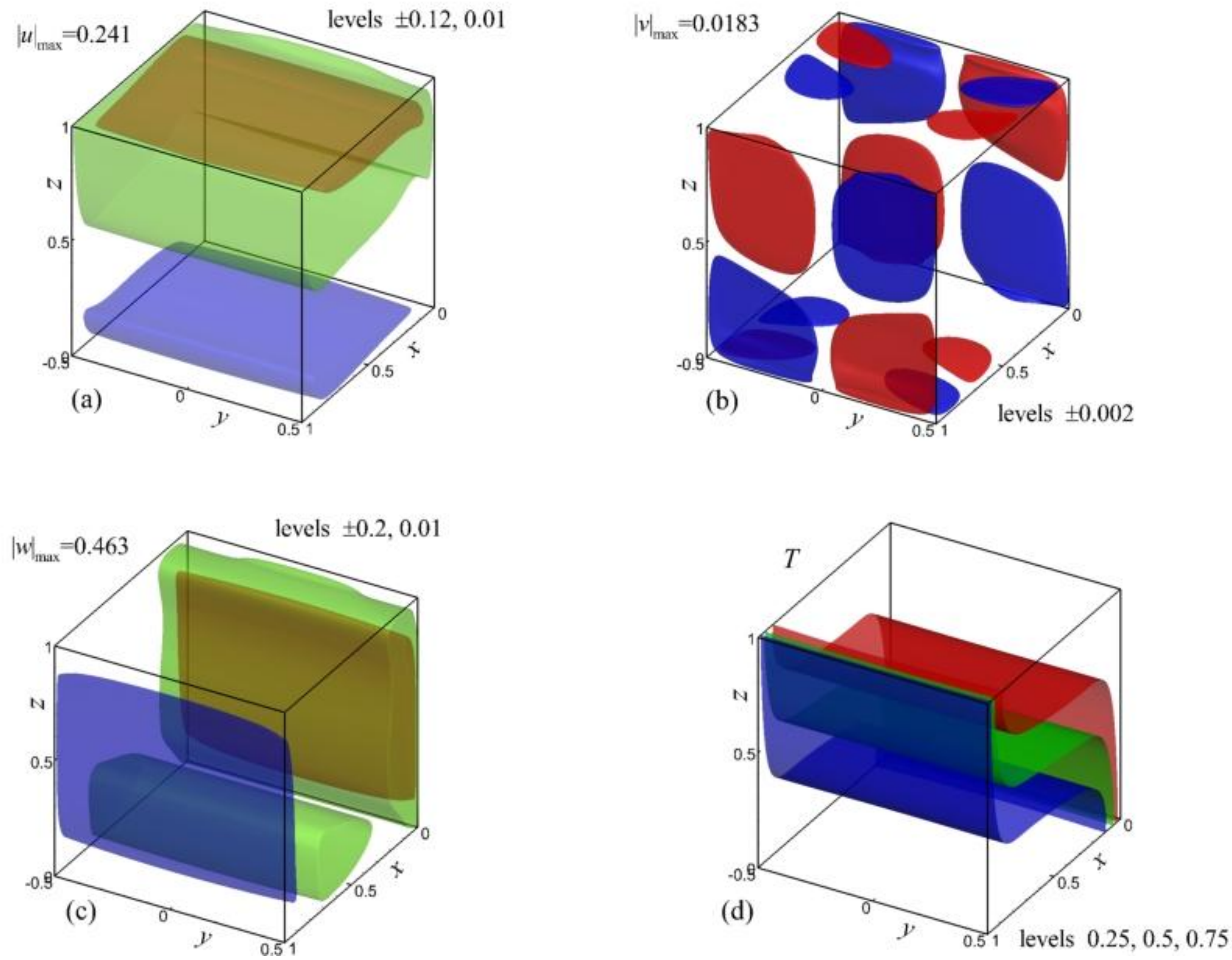

**Figure 8.** Isosurfaces of three velocity components and temperature. Cubic cavity with electrically insulated boundaries. $A = W = 1, Pr = 0.054,\ \ Ra = GrPr = 10^6, Ha = 1000, s = 10$.

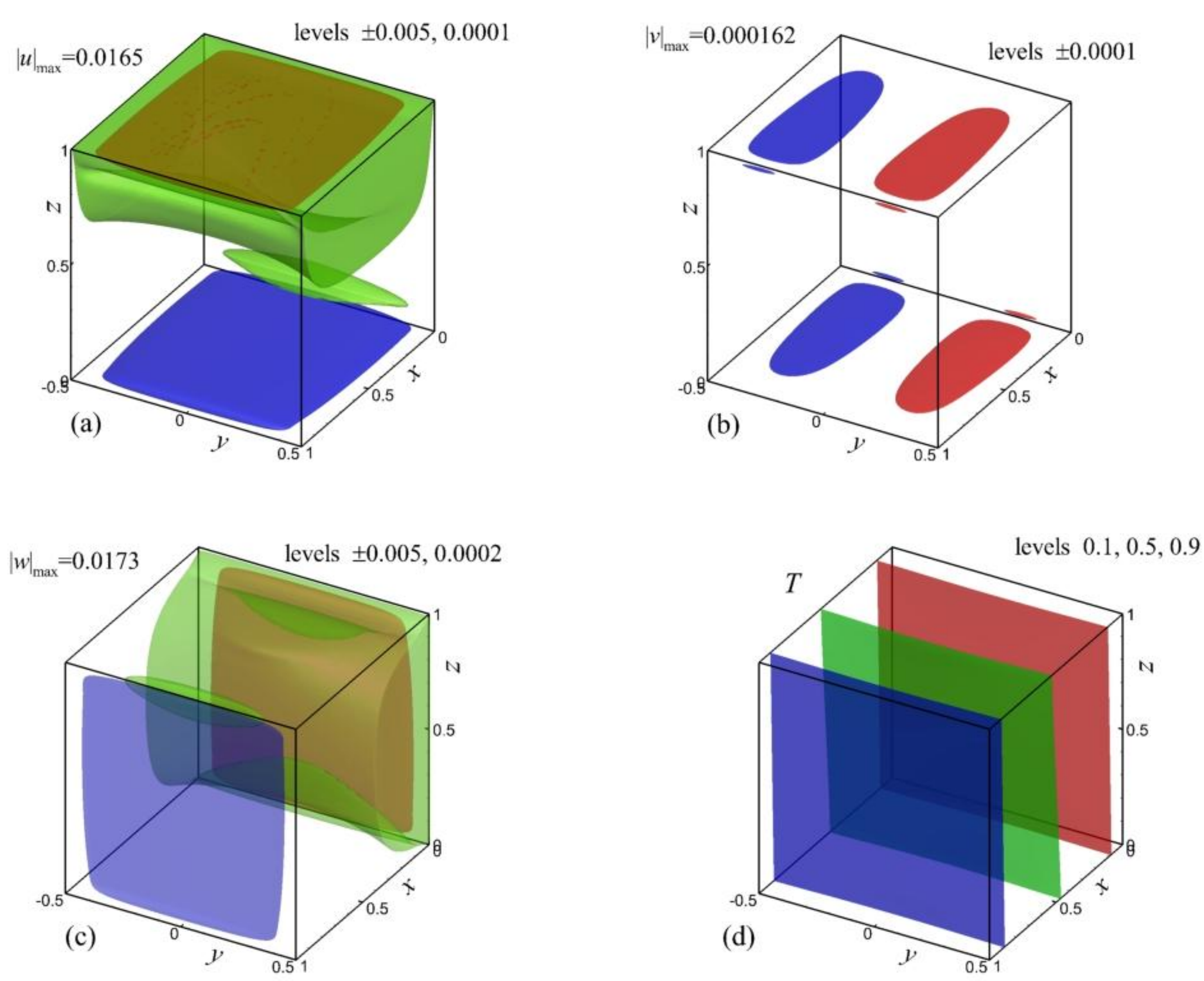

**Figure 9.** Isosurfaces of three velocity components and temperature. Cubic cavity with electrically conducting boundaries. $A = W = 1, Pr = 0.054, \ \ Ra = GrPr = 10^6, Ha = 1000, s = 10$.

.

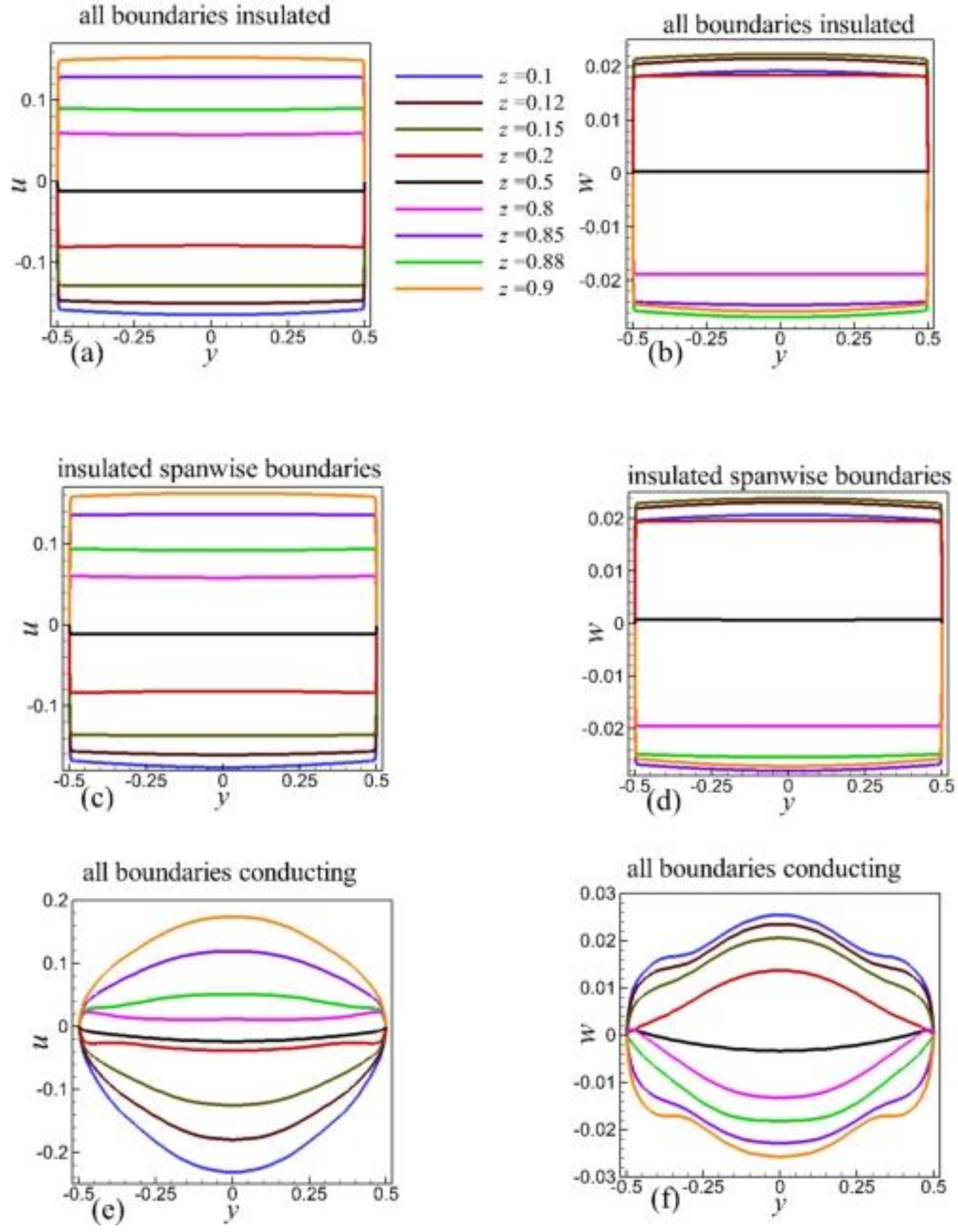

**Figure 10.** Profiles of the horizontal and vertical velocity components along the spanwise coordinate at $x = 0.5$ for three different boundary conditions for the electric potential. (a), (b) – all boundaries are electrically insulated, $s = 10$; (c), (d) – spanwise boundaries are electrically insulated, others are electrically conducting, $s = 10$; (e), (f) – all boundaries are electrically conducting, $s = 10$ . $A = W = 1, Pr = 0.054, \ \ Ra = GrPr = 10^6, Ha = 1000$.

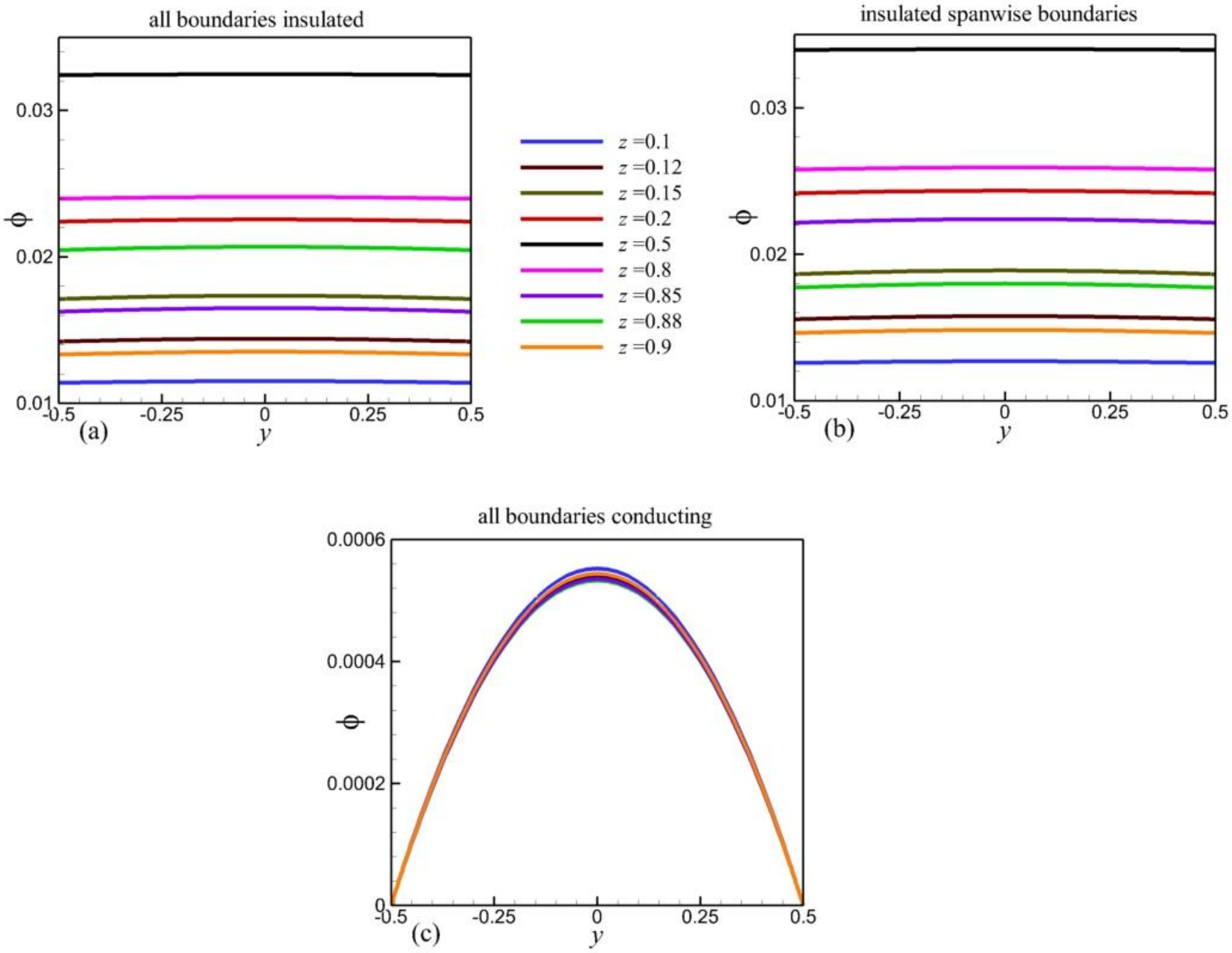

**Figure 11.** Potential profiles along the spanwise coordinate for three different boundary conditions for the electric potential. a) – all boundaries are electrically insulated, $s = 10$; (b) – spanwise boundaries are electrically insulated, others are electrically conducting, $s = 10$; (c) – all boundaries are electrically conducting. $A = W = 1, Pr = 0.054$, $Ra = GrPr = 10^6, Ha = 1000, s = 10$.

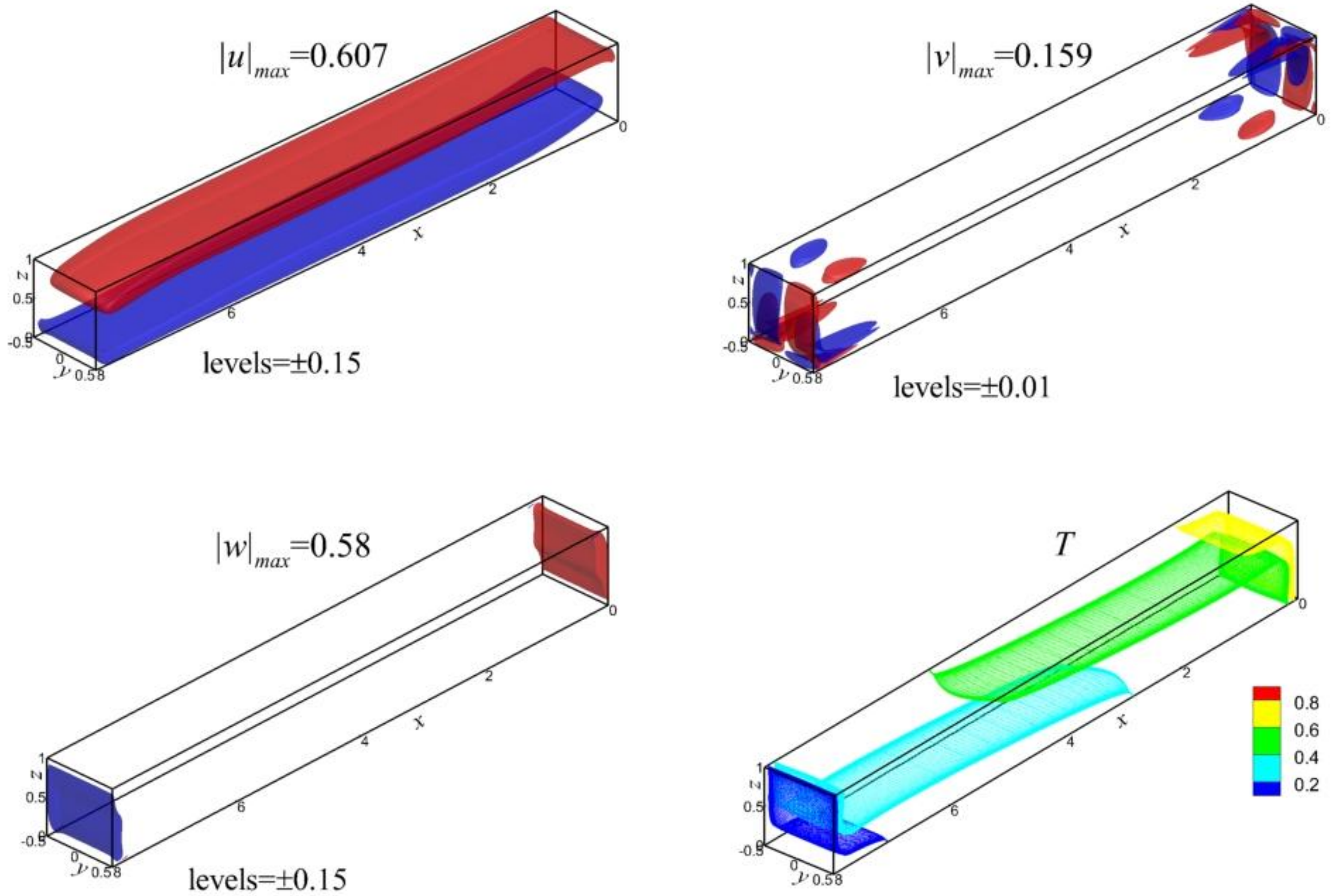

**Figure 12.** Isosurfaces of three velocity components and temperature. Horizontally elongated cavity with electrically insulated boundaries.

$A = 8, W = 1, Pr = 0.015, \; Gr = 8 \times 10^7, Ha = 100, s = 6.$

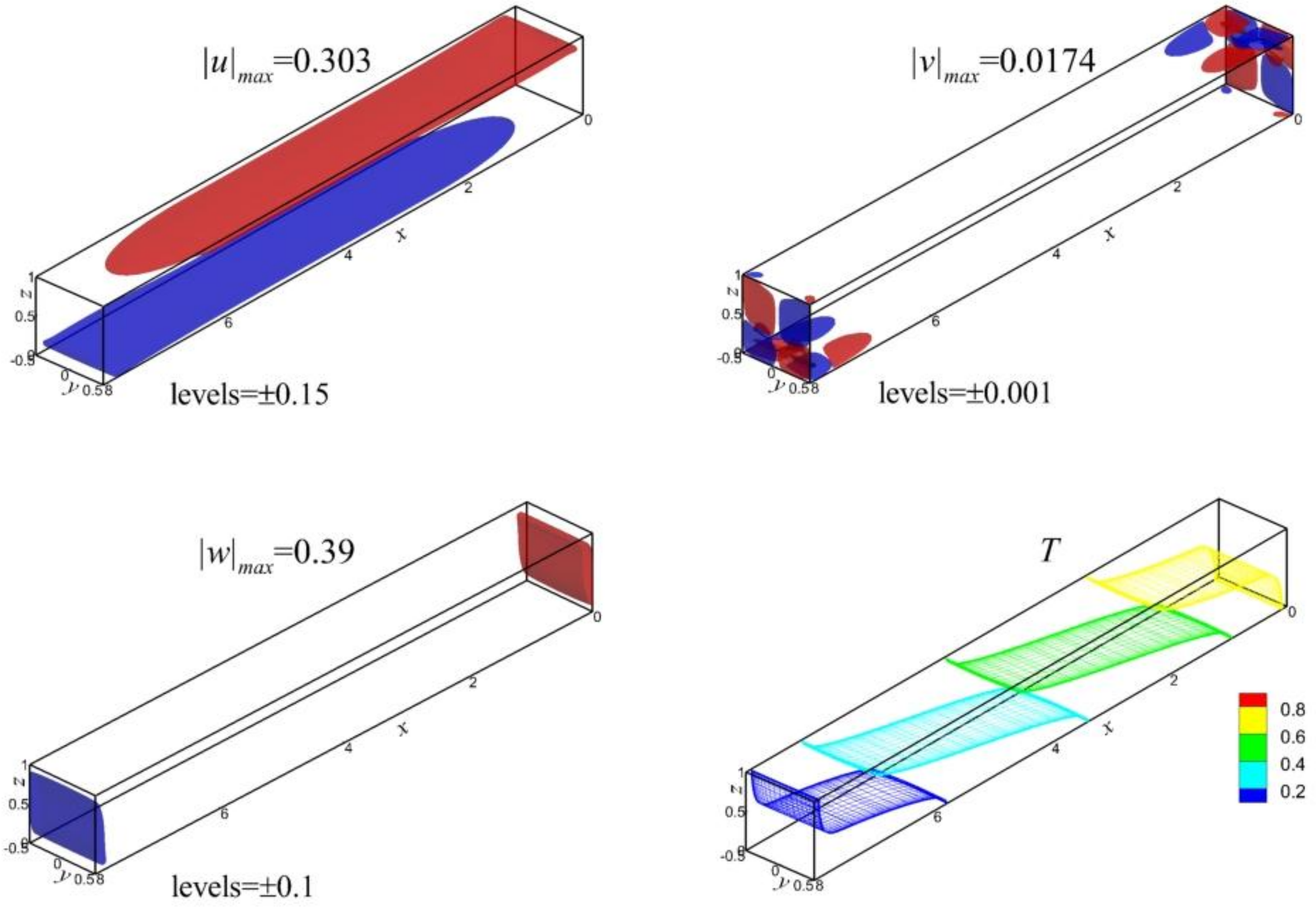

**Figure 13.** Isosurfaces of three velocity components and temperature. Horizontally elongated cavity with electrically insulated boundaries. $A = 8,\ W = 1,\ Pr = 0.015,\ \ Gr = 8 \times 10^7, Ha = 1000, s = 10.$

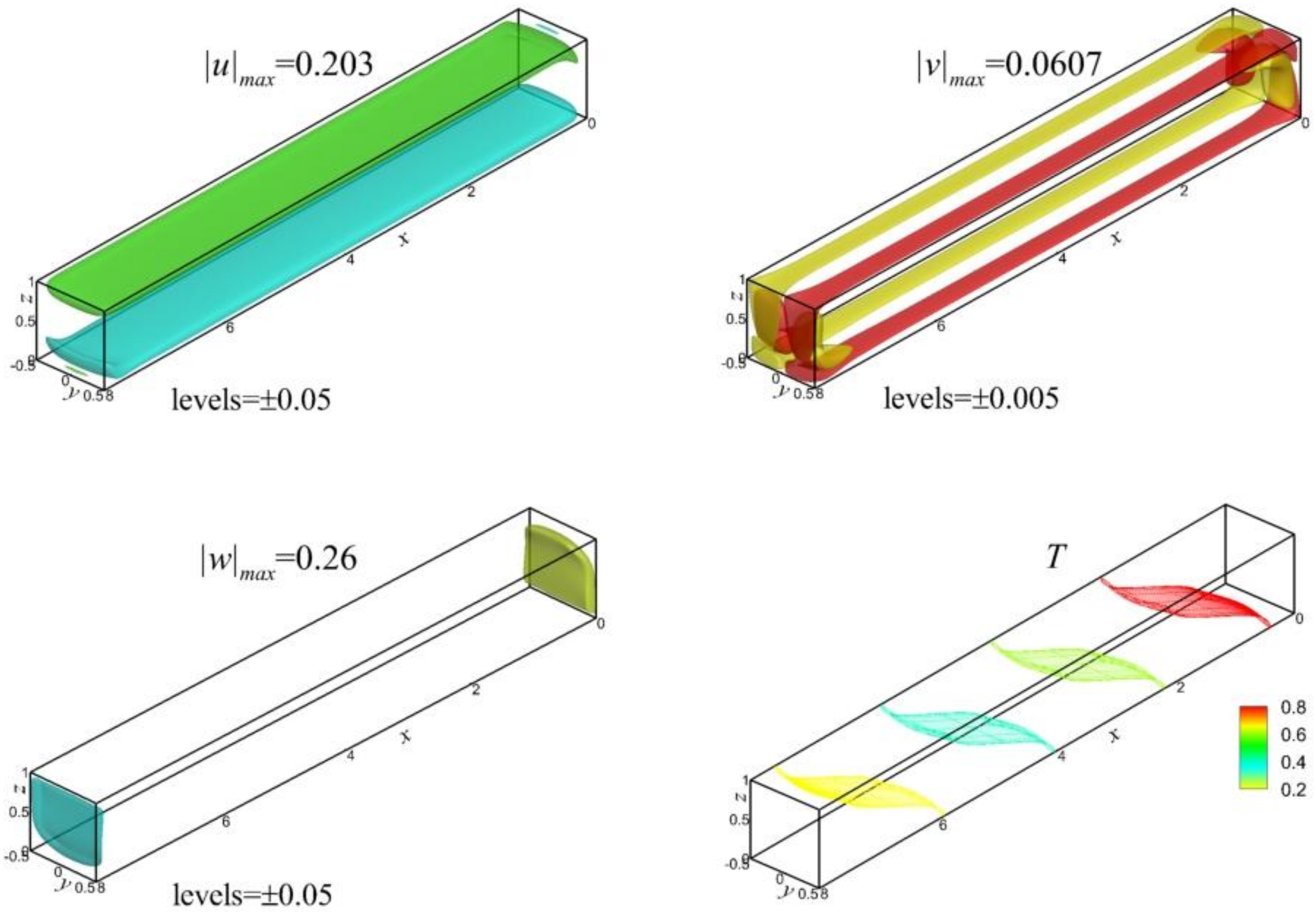

**Figure 14.** Isosurfaces of three velocity components and temperature. Horizontally elongated cavity with electrically conducting boundaries. $A = 8,\ W = 1,\ Pr = 0.015,\ \ Gr = 8 \times 10^7, Ha = 100, s = 6.$

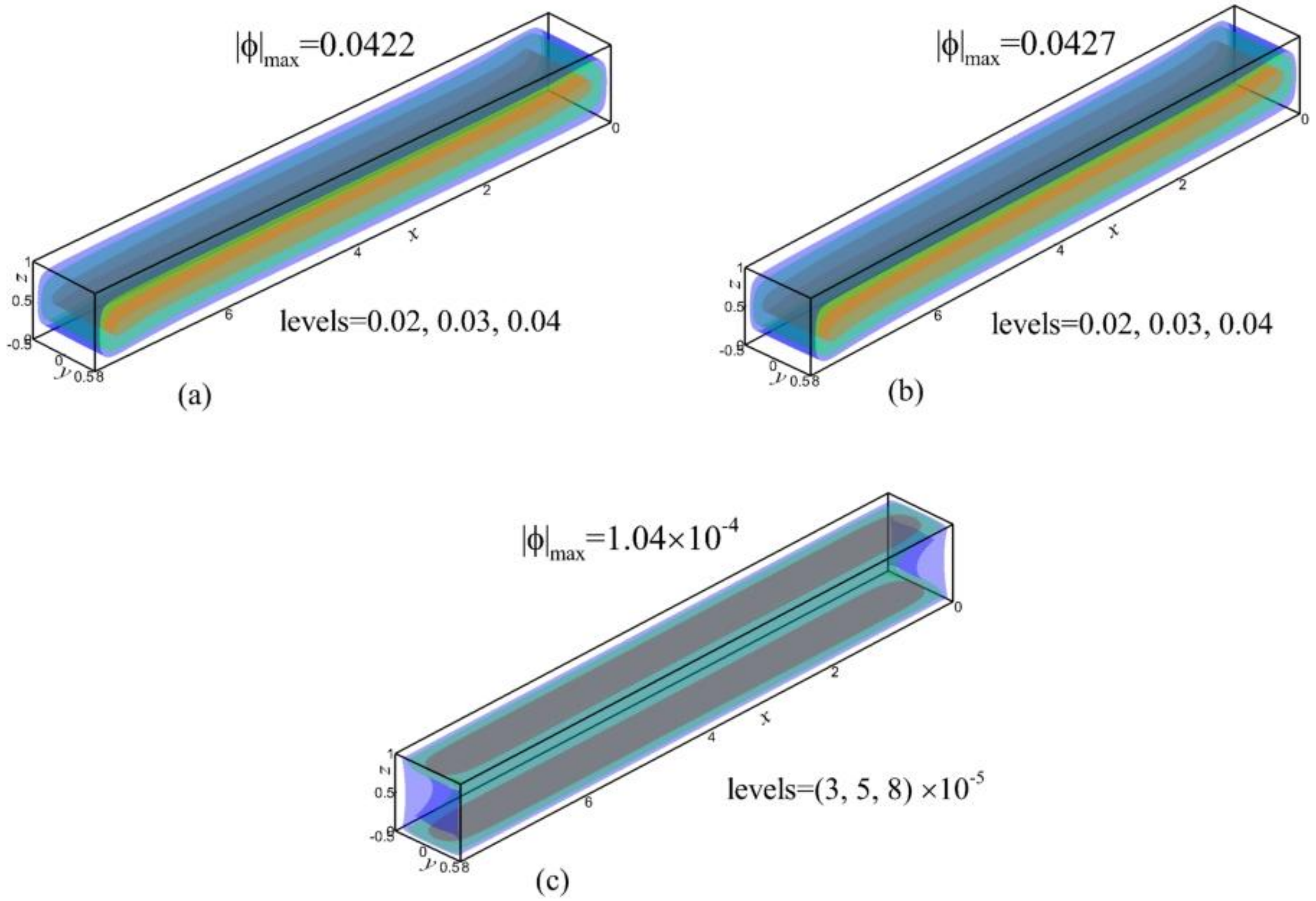

**Figure 15.** Equipotential surfaces for three different boundary conditions for the electric potential. (a) – all boundaries are electrically insulated; (b) – spanwise boundaries are electrically insulated, others are electrically conducting; (c) – all boundaries are electrically conducting. $A = 8,\ W = 1,\ Pr = 0.015,\ \ Gr = 8 \times 10^7, Ha = 1000, s = 6$.

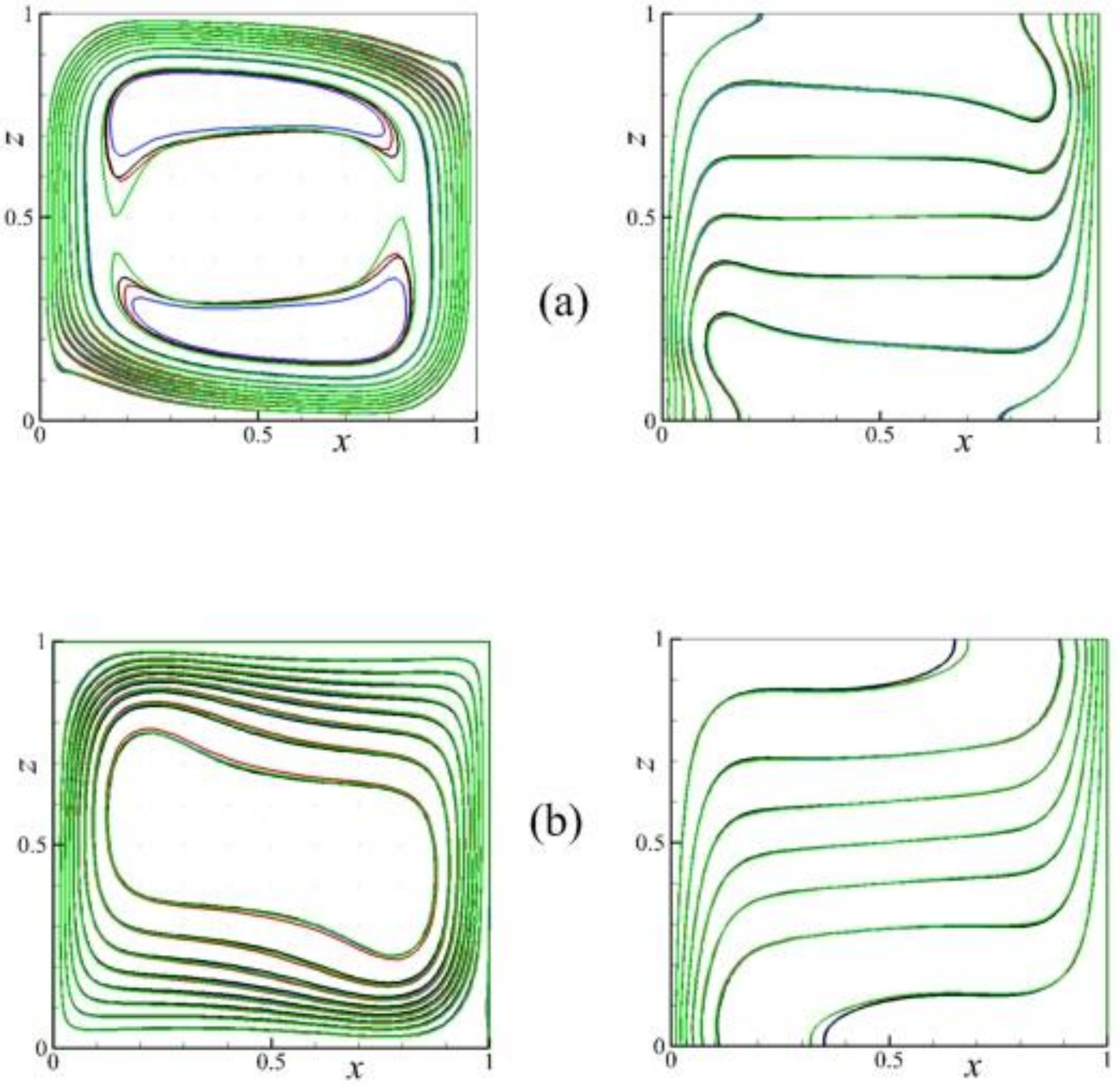

**Figure 16.** Streamlines (left) and isotherms (right) of $y$-averaged 3D and Q2D flows. $A = 1, Pr = 0.054,\ \ Ra = GrPr = 10^6$. Red – all boundaries insulated ($s = 10$), blue – spanwise boundaries insulated, others conducting ($s = 10$), both for $W = 1$. Green lines - all boundaries insulated, $W = 5$, ($s = 15$). Black lines – Q2D model, $s = 4$. (a) $Ha = 100$, the stream function maximum in different cases varies from 0.0466 to 0.0473. (b) $Ha = 1000$, the stream function maximum in different cases varies from 0.033 to 0.035. The streamlines are equally distributed between 0 and 0.05 for both values of $Ha$. The isotherms are equally distributed between 0 and 1.

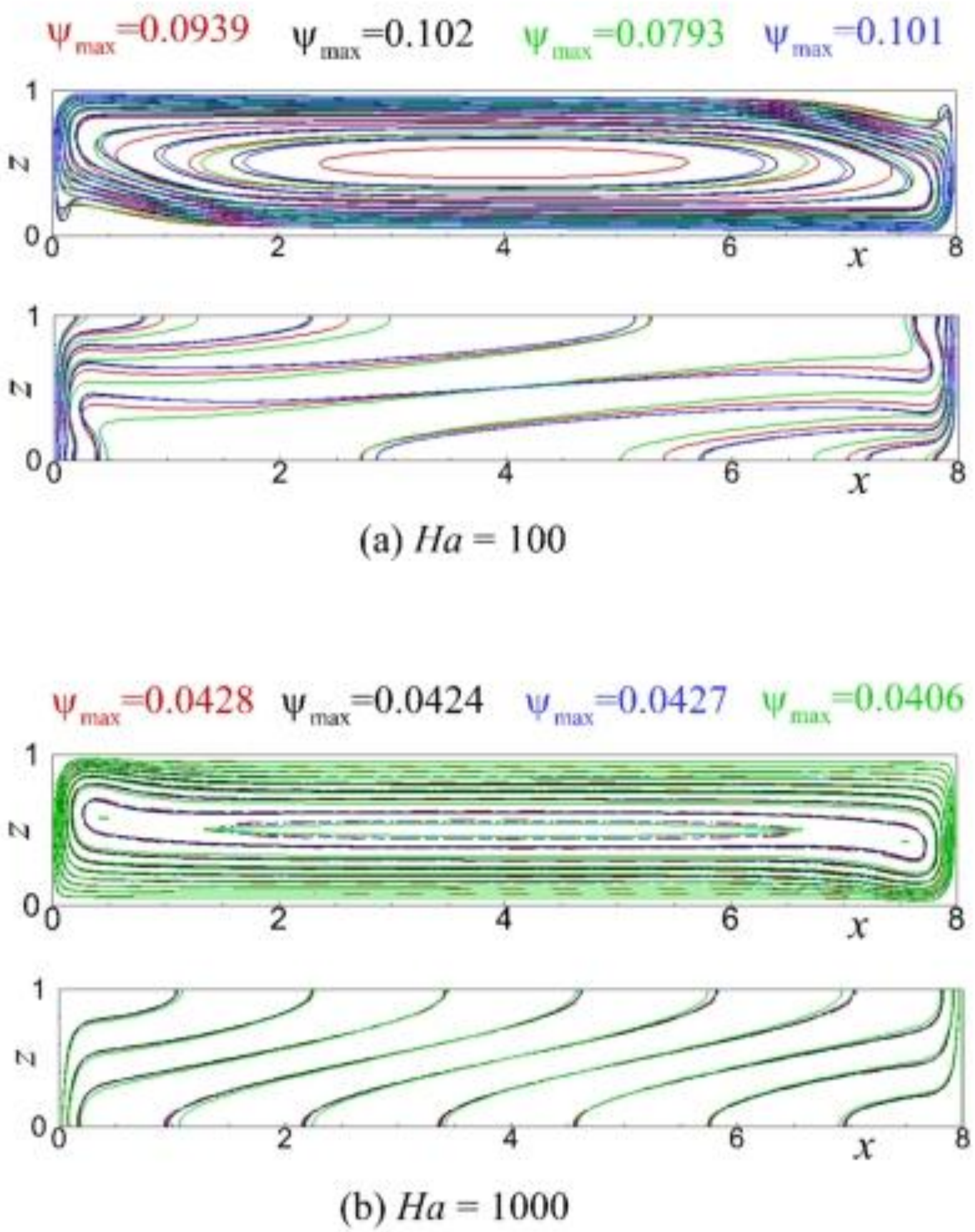

**Figure 17.** Streamlines (upper frames) and isotherms (lower frames) of $y$-averaged 3D and Q2D flows. $A = 8, Pr = 0.015,\ Gr = 8 \times 10^7$. Red – all boundaries insulated ($s = 10$), blue – spanwise boundaries insulated, others conducting, both for $W = 1$, ($s = 10$). Green lines - all boundaries insulated, $W = 10$, ($s = 13$ for $Ha = 100$ and $s = 16$ for $Ha = 1000$). Black lines – Q2D model ($s = 8$). (a) $Ha = 100$, (b) $Ha = 1000$. The streamlines are equally distributed between 0 and 0.1 in the frame (a), red, black, and blue streamlines are equally distributed between 0 and 0.04 in frame (b), green streamlines are equally distributed between 0 and 0.026. The isotherms are equally distributed between 0 and 1 in all frames.

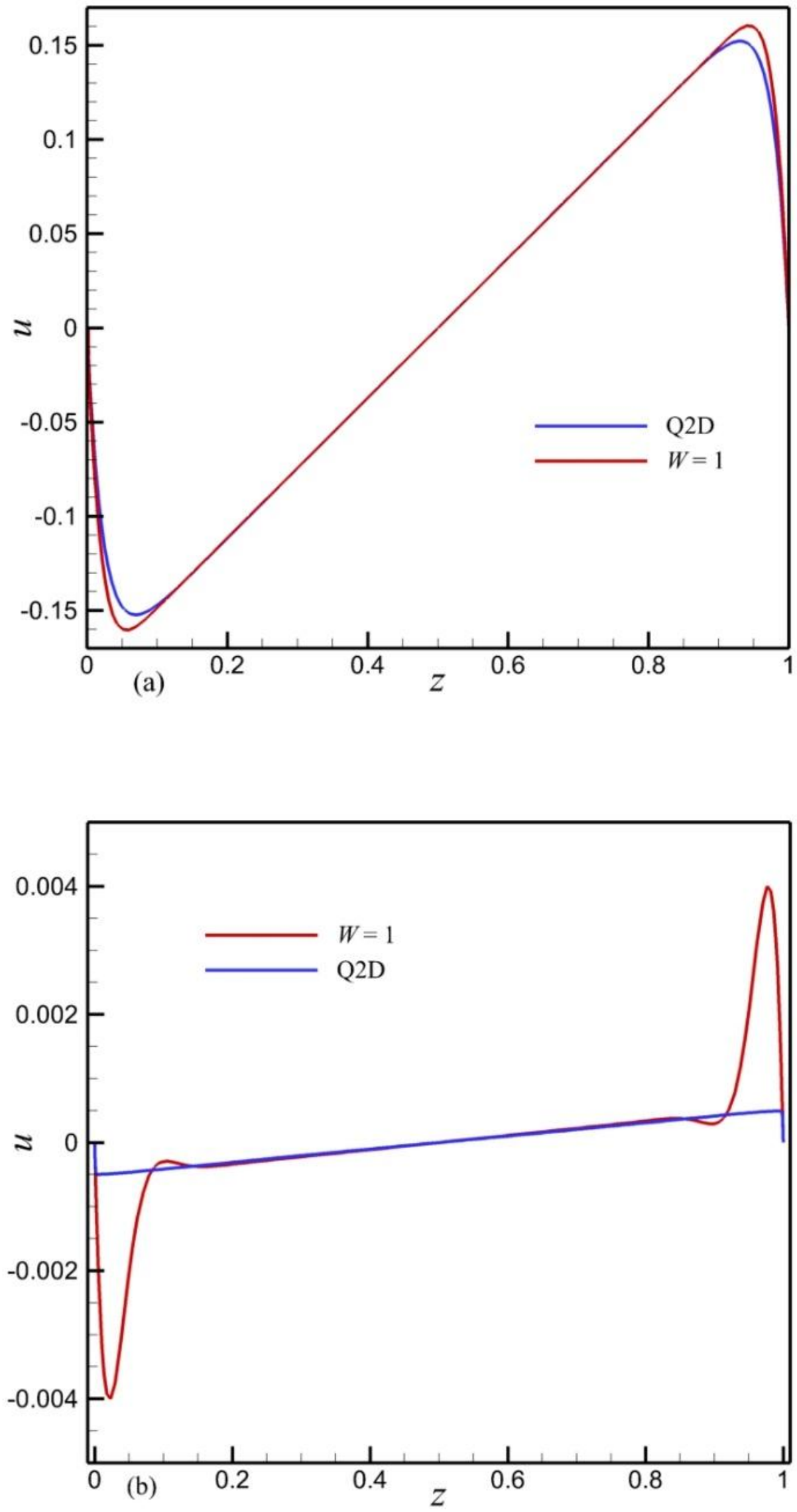

**Figure 18.** Comparison of the horizontal velocity profiles at $x = 0.5$ and $y = 0$ calculated by Q2D and fully 3D models. $A = 8, W = 1, Pr = 0.015,\ \ Gr = 8 \times 10^7, Ha = 1000, Hd = 2000$. (a) insulating boundaries, (b) conducting boundaries

.

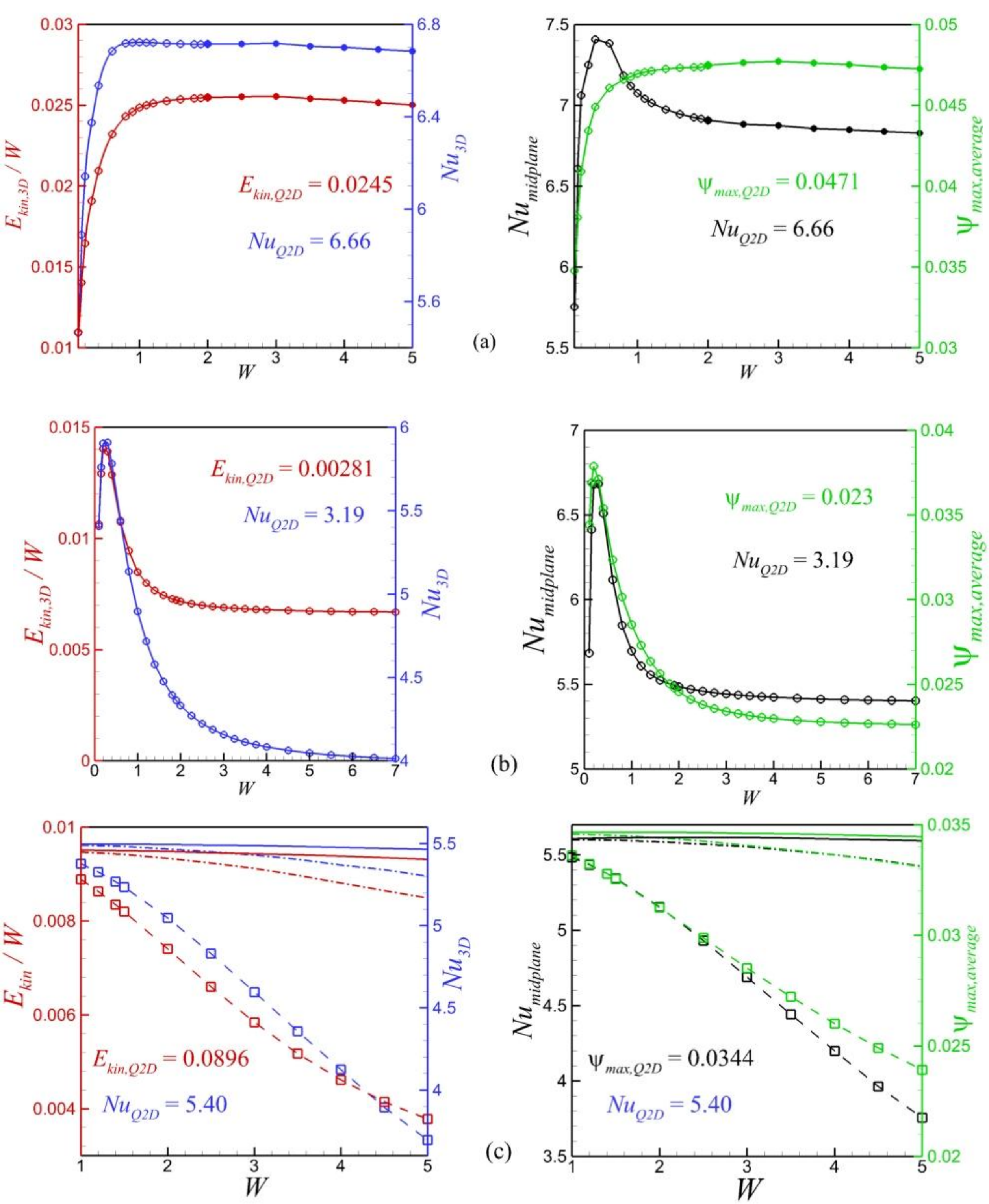

**Figure 19.** Dependence of the total kinetic energy and Nusselt number of fully 3D flow on the width ratio (left frames) and maximal value of the stream function of the *y*-averaged flow and the midplane Nusselt number (right frames). $A = 1, Pr = 0.054, \; Ra = GrPr = 10^6, Hd = 2Ha/W = 200$. (a) insulating boundaries, $Hd = 200$, (b) conducting boundaries $Hd = 200$, (c) insulating boundaries, $Hd = 2000$. Empty symbols correspond to steady flow, filled symbols – to the oscillatory ones. In frames (c) solid lines correspond to the stretching parameter $s = 15$, dash-and-dot lines to $s = 13$, and dash lines to $s = 10$.

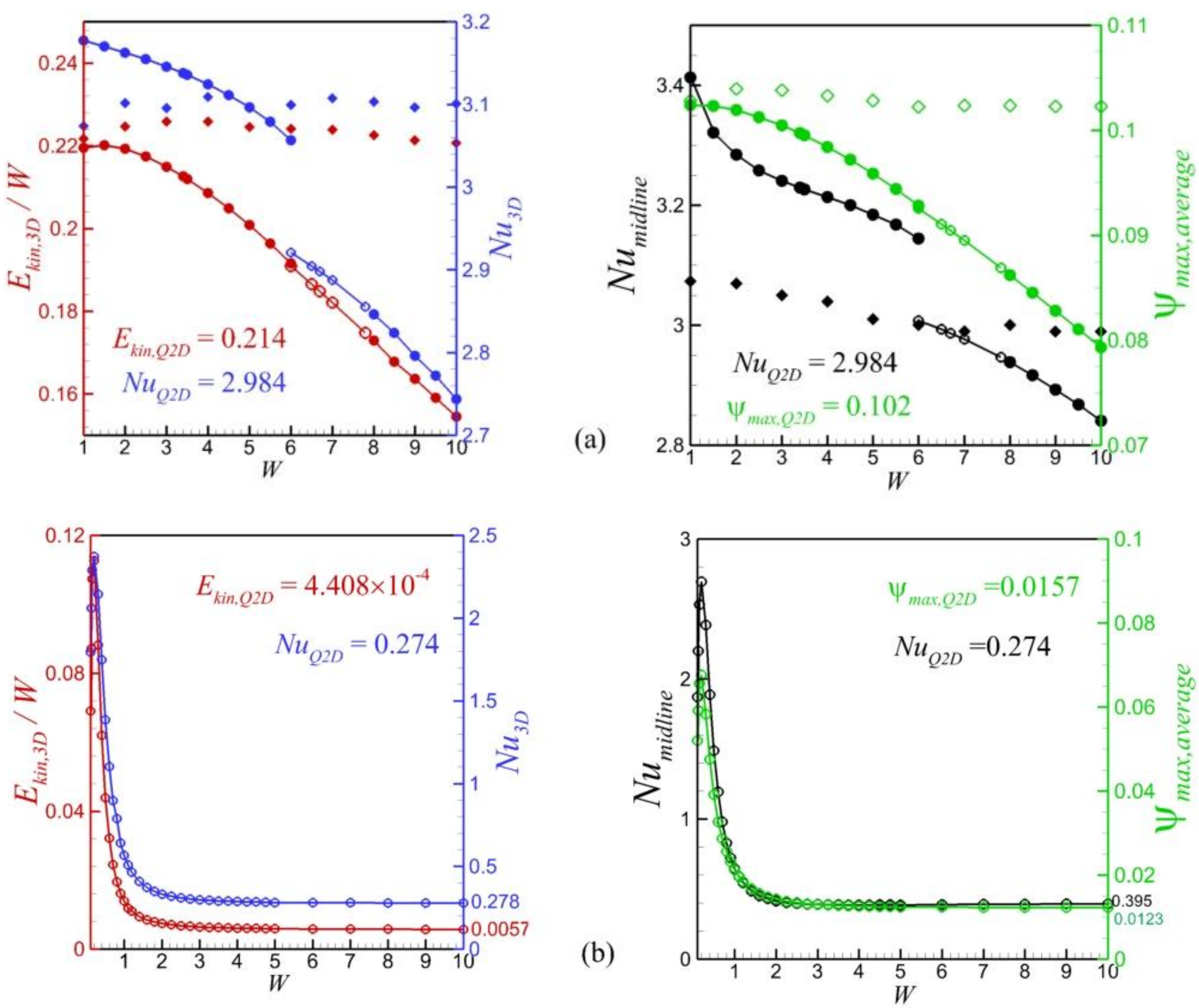

**Figure 20.** Dependence of the total kinetic energy and Nusselt number of fully 3D flow on the width ratio (left frames) and maximal value of the stream function and Nusselt number of the $y$-averaged flow (right frames). $A = 8, Pr = 0.015$, $Gr = 8 \times 10^7, Hd = 2Ha/W = 200$. (a) insulating boundaries, (b) conducting boundaries. Empty symbols correspond to steady flows, filled symbols – to oscillatory ones. In frames (a) the stretching parameter is $s = 10$ for circles and $s = 13$ for diamonds. In frames (b) the stretching parameter is $s = 6$.

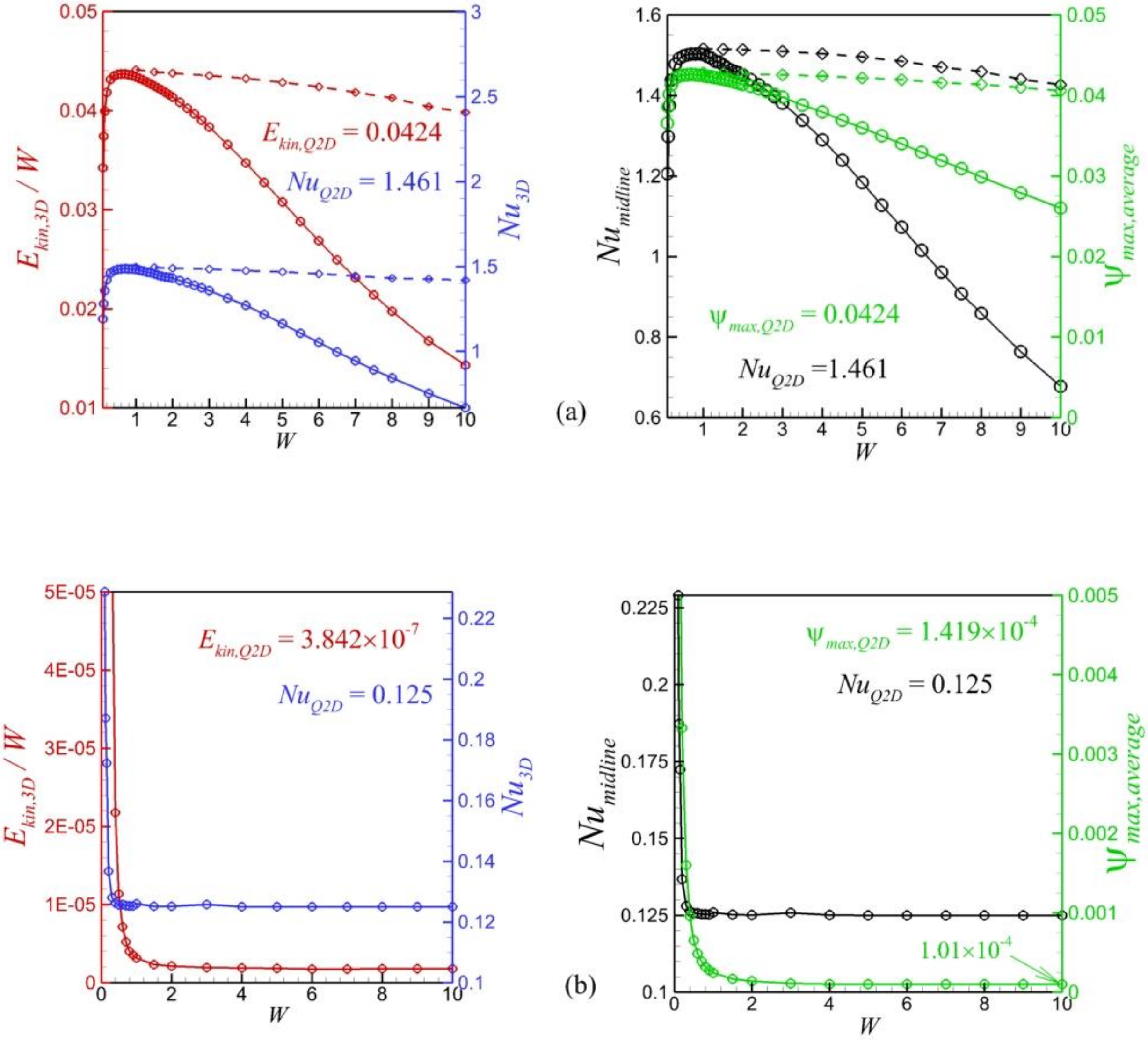

**Figure 21.** Dependence of the total kinetic energy and Nusselt number of fully 3D flow on the width ratio (left frames) and maximal value of the stream function and Nusselt number of the $y$-averaged flow (right frames). $A = 8, Pr = 0.015,\ Gr = 8 \times 10^7, Hd = 2Ha/W = 2000$. (a) insulating boundaries, the stretching parameter is $s = 13$ for circles and $s = 16$ for diamonds (b) conducting boundaries, $s = 13$.

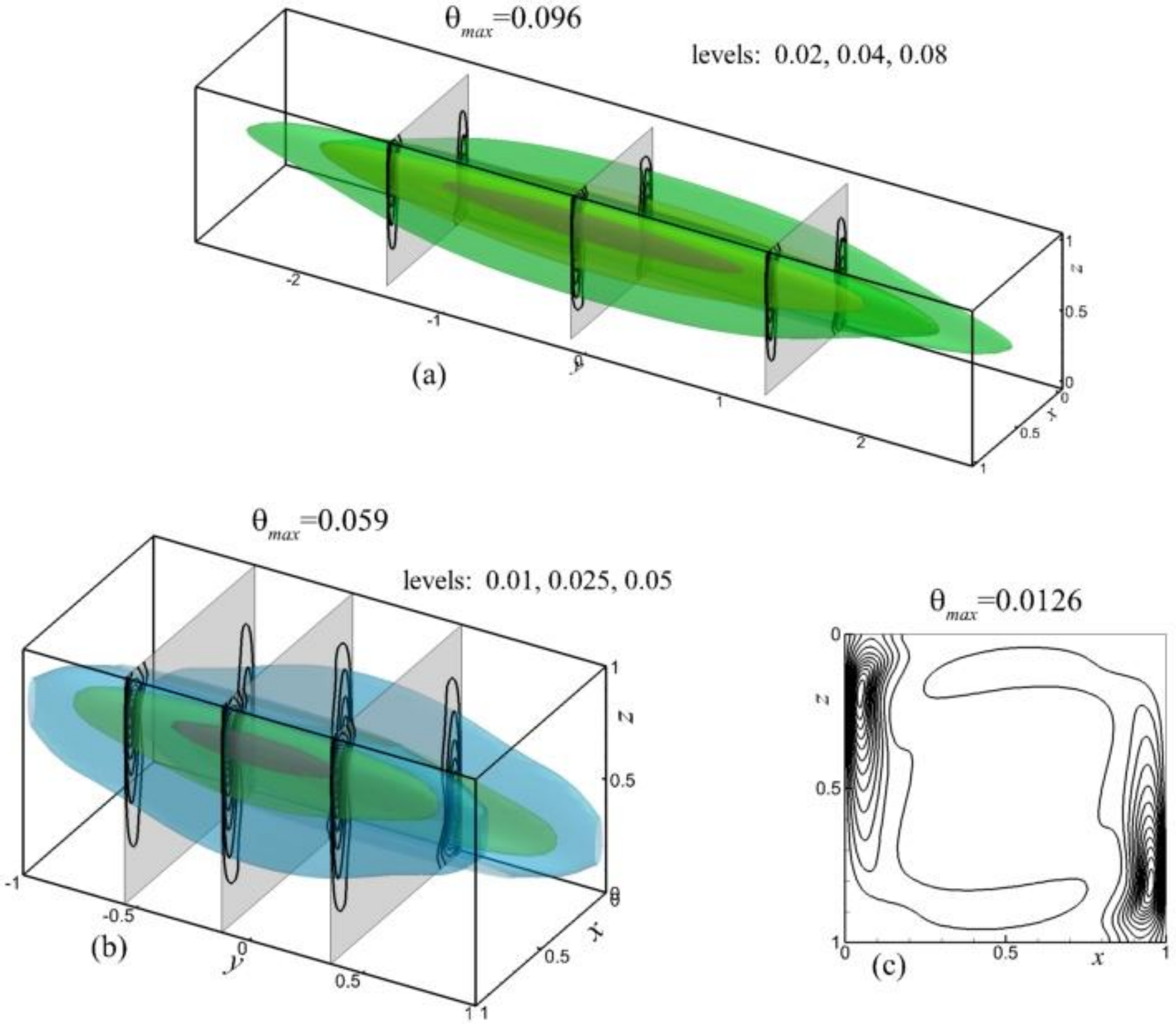

**Figure 22.** (a), (b) Amplitudes of temperature oscillations of the fully 3D flow at $A = 1, Pr = 0.054,\ \ Ra = GrPr = 10^6, Ha = 100$, insulating spanwise boundaries, for $W = 5$ and 2, respectively. (c) Amplitude of the most unstable temperature disturbance of the Q2D model for $A = 1, Pr = 0.054,\ \ Ra = GrPr = 1.08 \times 10^6, Hd = 200$.